\documentclass{article}
\usepackage{float}
\usepackage{subfig}
\usepackage{graphicx}
\usepackage{bibtopic}
\usepackage{amsmath}
\usepackage{attachfile}
\usepackage[
top    = 0.5in,
bottom = 0.50in,
left   = 0.37in,
right  = 0.37in]{geometry}
\usepackage{setspace}
\usepackage{chapterbib}
\usepackage{cite}
\usepackage{breqn}
\usepackage{mathtools}
\usepackage{empheq}
\usepackage{lipsum}
\usepackage{bm}
\usepackage{mathtools, cuted}
\usepackage{hyperref}
\usepackage{multirow}
\usepackage{lipsum}
\usepackage{multicol}
\usepackage{enumerate}

\usepackage[T1]{fontenc}
\usepackage[utf8]{inputenc}
\usepackage[T1]{fontenc}
\usepackage{titlesec}

\usepackage{fancyhdr}

\makeatletter
\newcommand{\chapterauthor}[1]{%
  {\parindent0pt\vspace*{-25pt}%
  \linespread{1.1}\large\scshape#1%
  \par\nobreak\vspace*{35pt}}
  \@afterheading%
}
\makeatother

\title{\bf Optimizing the performance of an Astigmatic Defocus Sensor and its implementation using Computer Generated Hologram}
\author{{\Large \bf Santanu Konwar}\\
Abhayapuri College, Abhayapuri, Bongaigaon, Assam, India\\
Email: k.santanu@alumni.iitg.ac.in}
\date{\today}

\begin{document}
\maketitle

\begin{multicols}{2}

\begin{abstract}
Astigmatic Defocus Sensor is a very simple and easy technique of measuring the amount of defocus (deviation from the focal plane) present in an optical beam. The approach use a cylindrical lens to focus the beam and the measure of the intensity at the focal plane is used to find out the amount of defocus present in the beam. In this paper we provide a theoretical discussion, on the Astigmatic Defocus Sensor, based on Fourier optics. The theoretical discussion is validated through simulation results. Two ways to optimize the performance of the astigmatic defocus sensor were also established. Further, the process of implementing the Astigmatic Defocus Sensor using computer generated hologram is presented.
\end{abstract}

\section{Introduction}
An optical beam with astigmatism can be sharply focussed at two different planes perpendicular to the propagation direction. The beam when focused on one of the plane shows a intensity distribution of line shaped directed along a certain direction, on the other hand the intensity distribution focussed on the other plane is line shaped oriented in a direction rotated by $90^o$ degree to that of the previous one. The focal plane of the beam free from astigmatism lies in between these two planes, where the intensity distribution will appear as a cross due to the effect of astigmatism \cite{born2013principles}. This unique property of change in orientation of the intensity distribution of the beam in presence of astigmatism can be used for the measurement of focussing error or the amount of defocus present in an optical beam \cite{cohen1984automatic, mansuripur1987analysis}. The astigmatic lens approach is a very simple technique that gives a direct measure of the defocus present in the beam. The technique being free from any intensive computation, provides a very fast and effective measure of the focussing error and can be used for auto focusing an optical beam \cite{cohen1984automatic, hsu2009development}, dynamic focus control \cite{kazasidis2020sensor}, thickness measurement \cite{liu2008application}, beam tracking \cite{mansuripur1987analysis}, etc.\\

The defocus measurement can be carried out using a single astigmatic lens system \cite{cohen1984automatic, paterson2000hybrid} or a double astigmatic lens system \cite{mansuripur1987analysis, bai2018focusing}. However, in this paper we deal with the single astigmatic lens approach. We present an elaborate theoretical discussion on the sensor followed by simulation validation. The theoretical discussion is mainly based on Fourier optics\cite{goodman2005introduction} and the Zernike polynomials\cite{noll1976zernike}. We also deduce an expression for the sensitivity and discuss on the cross sensitivity of of the sensor. Simulations are carried out to test the validity of the theoretical findings. Further, two ways of optimizing the performance of the astigmatic defocus sensor are also established. Later, the implementation of the Astigmatic Defocus Sensor using computer generated hologram is presented.

\section{Theory}
\begin{figure}[H]
\centering
\includegraphics[scale=0.4]{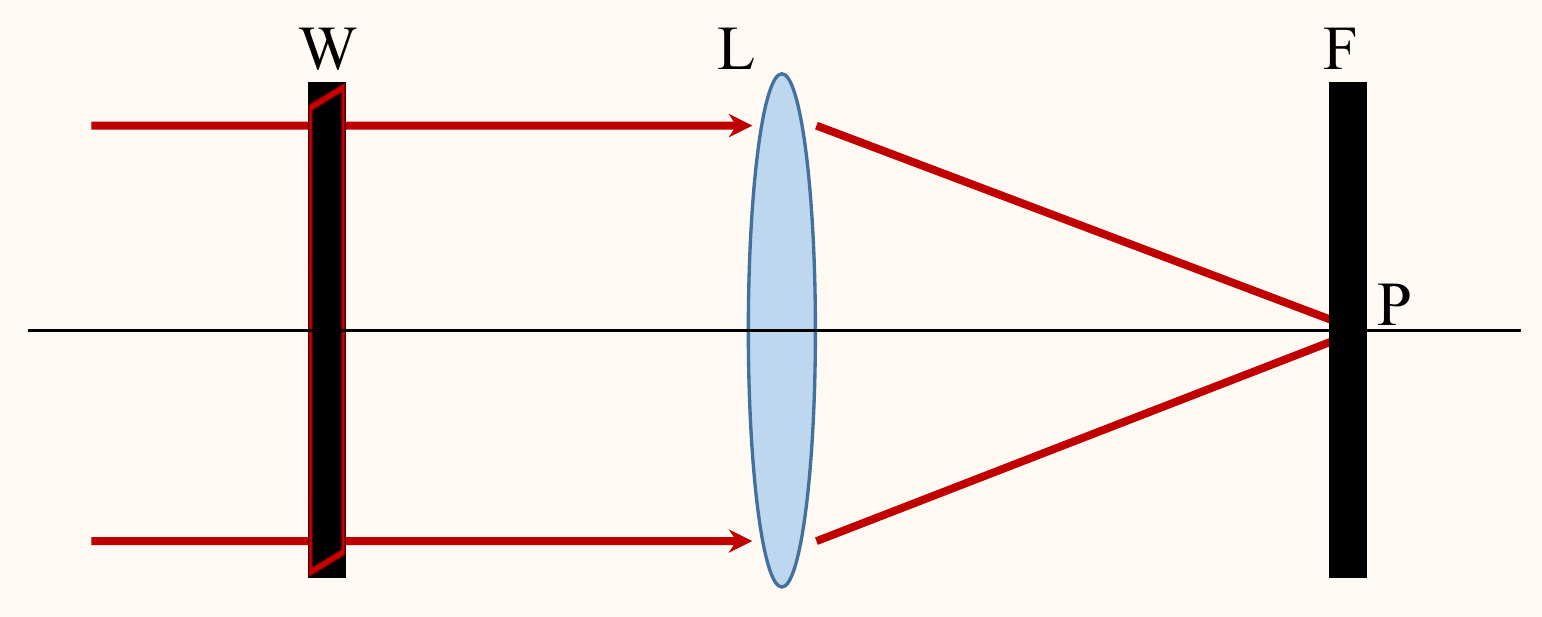}
\caption{Illustration of a monochromatic collimated beam with a wavefront $W$ passing through a converging lens $L$ and focussed at a point $P$ on the detector $F$}
\label{fig1}
\end{figure}

To begin with, let us consider that a monochromatic collimated beam with a certain phase $\psi (r,\theta)$ is passing through a converging lens of focal length $f$ as shown in Fig. \ref{fig1}. 

The amplitude distribution immediately after the lens $L$ will be

\begin{equation}
U'_L(x,y) = U_L(x,y)circ(r)e^{-j\frac{k}{2f}(x^2+y^2)}
\label{eq:refname3}
\end{equation}

where, $U_L(x,y)$ is the amplitude distribution of the wavefront $W$ incident on the lens given by

\begin{equation}
U_L(x,y) = A e^{j\psi (r,\theta)}
\label{eq:refname2}
\end{equation}
and $circ(r)$ defines a circular aperture given by
\[
    circ(r)= 
\begin{cases}
    1,& \text{for}\; r\leq 1\\
    0,& \text{for}\; r > 1
\end{cases}
\]

The amplitude distribution at the back focal plane $F$ of the lens can be obtained from the Fresnel
diffraction integral which can be written as \cite{goodman2005introduction}.

\begin{eqnarray}
U(u,v) &=& \frac{e^{jkz}}{j \lambda z} e^{j\frac{k}{2z}(u^2+v^2)}\int^{\infty}_{-\infty} \int^{\infty}_{-\infty}[U(x,y)e^{j\frac{k}{2z}(x^2+y^2)}]\nonumber \\
&&e^{-j\frac{2\pi}{\lambda z}(xu+yv)}dxdy
\label{eq:refname1}
\end{eqnarray}

Now, by taking $z=f$ equation \ref{eq:refname1} can be rewritten as

\begin{eqnarray}
U_F(u,v) &=& \frac{e^{jkf}}{j \lambda f} e^{j\frac{k}{2f}(u^2+v^2)}\int^{\infty}_{-\infty} \int^{\infty}_{-\infty}[U'_L(x,y)e^{j\frac{k}{2f}(x^2+y^2)}]\nonumber \\
&&e^{-j\frac{2\pi}{\lambda f}(xu+yv)}dxdy
\label{eq:refname4}
\end{eqnarray}

where $U_F(u,v)$ is the amplitude distribution at the back focal plane of the lens. On removing the constant phase and further simplification, we get

\begin{eqnarray}
&&U_F(u,v)\nonumber \\ 
&=& \frac{e^{j\frac{k}{2f}(u^2+v^2)}}{j \lambda f} \int^{\infty}_{-\infty} \int^{\infty}_{-\infty}[U_L(x,y)circ(r)e^{j\frac{k}{2f}(x^2+y^2)} \nonumber \\
&&e^{j\frac{k}{2f}(x^2+y^2)}] e^{-j\frac{2\pi}{\lambda f}(xu+yv)}dxdy \nonumber \\
&=& \frac{e^{j\frac{k}{2f}(u^2+v^2)}}{j \lambda f} \int^{\infty}_{-\infty} \int^{\infty}_{-\infty}U_L(x,y)circ(r)e^{-j\frac{2\pi}{\lambda f}(xu+yv)}dxdy \nonumber \\
\label{eq:refname5}
\end{eqnarray}
 
The expression in equation \ref{eq:refname5} can be transformed into polar coordinates using the condition (detail derivation of the expression is available in Annexure 1)
 
\begin{eqnarray}
\frac{1}{\lambda f}(ux+vy)&=& \rho r (cos\phi cos\theta + sin\phi sin\theta)\nonumber \\
&=& \rho r cos(\phi - \theta)
\label{eq:refname6}
\end{eqnarray}
 
Thus, in polar coordinates, equation \ref{eq:refname5} can be written as

\begin{equation}
U_F(\rho,\phi)=\frac{e^{j\frac{k}{2f}\rho^2}}{j \lambda f} \int^{2\pi}_{0} \int^{\infty}_{0}U_L(r,\theta)e^{-j2\pi\rho r cos(\phi - \theta)}rdrd\theta 
\label{eq:refname7}
\end{equation}

where, $\rho^2 = u^2+v^2$ and

\begin{equation}
U_L(r,\theta)=Ae^{j\psi(r,\theta)}
\label{eq:refname8}
\end{equation}

The term $circ(r)$ is omitted keeping in view that the co-ordinates taken now are polar coordinates for a circular window. The intensity distribution at the back focal plane of the lens can thus be written as

\begin{eqnarray}
I_F(\rho,\phi)&=&\left[\frac{e^{j\frac{k}{2f}\rho^2}}{j \lambda f} \int^{2\pi}_{0} \int^{\infty}_{0}U_L(r,\theta)e^{-j2\pi\rho r cos(\phi - \theta)}rdrd\theta \right]\nonumber\\
&\times &\left[\frac{e^{-j\frac{k}{2f}\rho^2}}{-j \lambda f} \int^{2\pi}_{0} \int^{\infty}_{0}U_L(r,\theta)e^{-j2\pi\rho r cos(\phi - \theta)}rdrd\theta \right]\nonumber\\
&=&\frac{1}{\lambda ^2 f^2}\left|\int^{2\pi}_{0} \int^{\infty}_{0}U_L(r,\theta)e^{-j2\pi\rho r cos(\phi - \theta)}rdrd\theta \right|^2
\label{eq:refname9}
\end{eqnarray}

The phase profile of the incident beam $\psi(r,\theta)$ can be represented in terms of Zernike modes as \cite{noll1976zernike}. 

\begin{equation}
\psi(r,\theta)=\sum^{\infty}_{k=0} a_k Z_k (r,\theta)
\label{eq:refname10}
\end{equation}

where $Z_k (r,\theta)$ are Zernike modes that can represent the optical aberrations modes correctly. The Zernike modes can be represented as \cite{noll1976zernike}

\begin{equation}
\begin{array}{l}
{\left. \begin{array}{l}
Z_{even\, k}=\sqrt{n+1} R_n^m (r) \sqrt{2} \cos(m\theta)\\
Z_{odd\, k}=\sqrt{n+1} R_n^m (r) \sqrt{2} \sin(m\theta)
\end{array} \right\}\,\mbox{for m$\ne$0}}\\
Z_{k}=\sqrt{n+1} R_n^m (r)\,\mbox{\; \; for m=0}
\end{array}
\label{eq:refname11}
\end{equation}

and

\begin{equation}
R_n^m(r)=\sum_{q=0}^{\frac{(n-m)}{2}} \frac{(-1)^q (n-q)!}{q! \left[\frac{n+m}{2}-q\right]!\left[\frac{n-m}{2}-q\right]!} r^{n-2q}
\end{equation}

Few low order Zernike polynomials are shown in table \ref{tab:zernike} below.

\begin{table}[H]
\centering
\caption{\bf Zernike polynomials representing few low order aberrations}
\begin{tabular}{ccc}
\hline
index ($j$) & $Z_j (r, \theta)$ & common aberration \\
\hline
$k = 4$ & $\sqrt3(2r^{2}-1$ & Defocus \\
$k = 5$ & $\sqrt6r^2sin{2\theta}$ & Astigmatism $\pm 45^o$ \\
$k = 6$ & $\sqrt6r^2cos{2\theta}$ & Astigmatism $0^o$ \\
$k = 7$ & $\sqrt8(3r^{2}-2r)sin\theta$ & y coma \\
$k = 8$ & $\sqrt8(3r^{2}-2r)cos\theta$ & x coma \\
$k = 9$ & $\sqrt8r^{3}sin{3\theta}$ & Trefoil y \\
$k = 10$ & $\sqrt8r^{3}cos{3\theta}$ & Trefoil x \\
$k = 11$ & $\sqrt5(6r^{4}-6r^{2}+1)$ & Spherical aberration \\
\hline
\end{tabular}
  \label{tab:zernike}
\end{table}

\subsection{Astigmatic Defocus Sensor}
If a beam with a plane wavefront passes through an astigmatic lens rotated at $45^o$, i.e. the lens $L$ of Fig \ref{fig1} is replaced by an astigmatic lens, the amplitude distribution immediately after the lens will be

\begin{equation}
U_L(r,\theta)=A e^{j a Z_5 (r,\theta)}
\label{eq:refname12}
\end{equation}

where $Z_5 (r,\theta)$ is the Zernike mode that represent astigmatism with $45^o$ tilt and $a_5$ is taken as $a$ for simplicity. If the Zernike modes are defined over an unit circular aperture, the intensity at the detector will be

\begin{equation}
I(\rho,\phi)=\frac{A}{\lambda ^2 f^2}\left|\int^{2\pi}_{0} \int^{1}_{0}e^{j a Z_5 (r,\theta)}e^{-j2\pi\rho r cos(\phi - \theta)}rdrd\theta \right|^2
\label{eq:refname13}
\end{equation}

For a finite size detector of radius $\rho_p$ defined in Fourier coordinate\cite{goodman2005introduction}, the intensity recorded by the detector can be obtained by taking the integration of the whole detector area

\begin{equation}
I_T(\rho,\phi)=\int^{2\pi}_{0} \int^{\rho_p}_{0}I(\rho,\phi)\rho d\rho d\phi
\label{eq:refname14}
\end{equation}

\begin{figure}[H]
\centering
\includegraphics[scale=0.5]{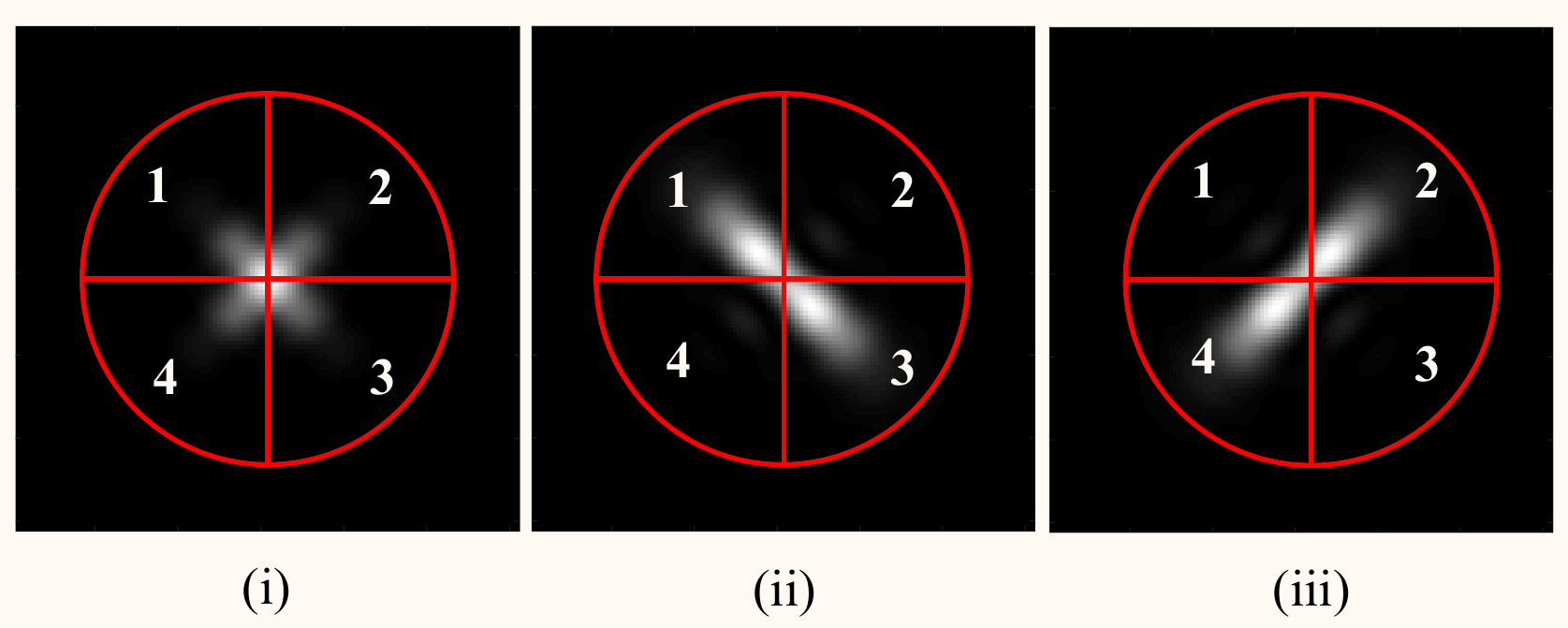}
\caption{Illustration of the intensity distribution at the detector plane with $a=1$ radian and (i) $b=0$ radian, (ii) $b=1$ radian and (iii) $b=-1$ radian}
\label{fig2}
\end{figure}

If we divide the detector aperture into four equal parts as shown in figure \ref{fig2}, the intensities at each sub-aperture will be

\begin{eqnarray}
I_1(\rho,\phi)&=&\int^{\frac{\pi}{2}}_{0} \int^{\rho_p}_{0}I(\rho,\phi)\rho d\rho d\phi \label{eq:refname15a}\\
I_2(\rho,\phi)&=&\int^{\pi}_{\frac{\pi}{2}} \int^{\rho_p}_{0}I(\rho,\phi)\rho d\rho d\phi\\
I_3(\rho,\phi)&=&\int^{\frac{3\pi}{2}}_{\pi} \int^{\rho_p}_{0}I(\rho,\phi)\rho d\rho d\phi \\
I_4(\rho,\phi)&=&\int^{2\pi}_{\frac{3\pi}{2}} \int^{\rho_p}_{0}I(\rho,\phi)\rho d\rho d\phi
\label{eq:refname15}
\end{eqnarray}

Now, let us consider that the incident beam consist of certain amount (say $b$ radian) of defocus ($Z_4$) as shown in figure \ref{fig2}(ii - iii). As seen in the figure, the presence of defocus incorporated directionality to the intensity profile at the detector plane. Further the intensity profile changes the orientation for opposite polarity of defocus. The intensity at the detector plane now becomes

\begin{equation}
I(\rho,\phi)=\left|\int^{2\pi}_{0} \int^{1}_{0}e^{jaZ_5 (r,\theta)+jbZ_4 (r,\theta)}e^{-j2\pi\rho r cos(\phi - \theta)}rdrd\theta \right|^2
\label{eq:refname16}
\end{equation}

The constant term $\frac{A}{\lambda ^2 f^2}$ is taken as 1 to keep the expression simple. The amount of defocus present in the incident beam can be obtained from the record of the intensities of the different sub-apertures. The sensor output can thus be obtained from the expression \cite{paterson2000hybrid}

\begin{equation}
S=\frac{(I_1+I_3)-(I_2+I_4)}{(I_1+I_3)+(I_2+I_4)}
\label{eq:refname17}
\end{equation}

It is quite clear from figure \ref{fig2}(ii - iii) that there is symmetry between the diagonally opposite sub-apertures and the total intensities of the diagonally opposite sub-apertures are same. Thus, $I_1=I_3$ and $I_2=I_4$ in both the cases of $b=1$ and $b=-1$. Therefore the expression of sensor output can be simplified as

\begin{equation}
S=\frac{(I_1-I_2)}{(I_1+I_2)}
\label{eq:refname18}
\end{equation}

\subsection{Sensitivity of Astigmatic Defocus Sensor}
The sensitivity of a sensor is a measure of the effectiveness of the sensor while measuring small aberration. It can be calculated by differentiating the sensor output by the amplitude of the input aberration mode ($Z_4$ in this case) i.e. $b$, at the amplitude of the bias aberration mode ($Z_5$ in this case) $a=0$ \cite{paterson2000hybrid}. Therefore proceeding accordingly we first calculate $\frac{\partial S}{\partial b}$ as

\begin{eqnarray}
\frac{\partial S}{\partial b}&=&\frac{(I_1+I_2)\frac{\partial}{\partial b}(I_1-I_2)-(I_1-I_2)\frac{\partial}{\partial b}(I_1+I_2)}{(I_1+I_2)^2}\nonumber \\
&=&\frac{2I_2\frac{\partial}{\partial b}I_1-2I_1\frac{\partial}{\partial b}I_2}{(I_1+I_2)^2}
\label{eq:refname19}
\end{eqnarray}

And,

\begin{eqnarray}
&&\frac{\partial}{\partial b}\left\{ I(u,v)\right \}\nonumber \\
&=&\frac{\partial}{\partial b}\left|\int^{2\pi}_{0} \int^{1}_{0}e^{jaZ_5 (r,\theta)+jbZ_4 (r,\theta)}e^{-j2\pi\rho r cos(\phi - \theta)}rdrd\theta \right|^2\nonumber \\
&=&\frac{\partial}{\partial b}\left[\int^{2\pi}_{0} \int^{1}_{0}e^{j[aZ_5 (r,\theta)+bZ_4 (r,\theta)-2\pi\rho r cos(\phi - \theta)}rdrd\theta \right.\nonumber \\
&&\times \left.\int^{2\pi}_{0} \int^{1}_{0}e^{-j[aZ_5 (r,\theta)+bZ_4 (r,\theta)-2\pi\rho r cos(\phi - \theta)}rdrd\theta\right]\nonumber \\
&=&-j\int^{2\pi}_{0} \int^{1}_{0}e^{j[aZ_5 (r,\theta)+bZ_4 (r,\theta)-2\pi\rho r cos(\phi - \theta)}rdrd\theta \nonumber \\
&&\times \int^{2\pi}_{0} \int^{1}_{0}Z_4 (r,\theta)e^{-j[aZ_5 (r,\theta)+bZ_4 (r,\theta)-2\pi\rho r cos(\phi - \theta)}rdrd\theta \nonumber \\
&& + \: j\int^{2\pi}_{0} \int^{1}_{0}e^{-j[aZ_5 (r,\theta)+bZ_4 (r,\theta)-2\pi\rho r cos(\phi - \theta)}rdrd\theta \nonumber \\
&& \times \int^{2\pi}_{0} \int^{1}_{0}Z_4 (r,\theta)e^{j[aZ_5 (r,\theta)+bZ_4 (r,\theta)-2\pi\rho r cos(\phi - \theta)}rdrd\theta\nonumber \\
&&
\label{eq:refname20}
\end{eqnarray}

which gives

\begin{eqnarray}
&&\left.\frac{\partial}{\partial b}\left\{ I(u,v)\right \} \right|_{b=0} \nonumber \\
&=&-j\left[\int^{2\pi}_{0} \int^{1}_{0}e^{j[aZ_5 (r,\theta)-2\pi\rho r cos(\phi - \theta)}rdrd\theta \right. \nonumber \\
&&\times \int^{2\pi}_{0} \int^{1}_{0}Z_4 (r,\theta)e^{-j[aZ_5 (r,\theta)-2\pi\rho r cos(\phi - \theta)}rdrd\theta \nonumber \\
&& - \: \int^{2\pi}_{0} \int^{1}_{0}e^{-j[aZ_5 (r,\theta)-2\pi\rho r cos(\phi - \theta)}rdrd\theta \nonumber \\
&& \times \left.\int^{2\pi}_{0} \int^{1}_{0}Z_4 (r,\theta)e^{j[aZ_5 (r,\theta)-2\pi\rho r cos(\phi - \theta)}rdrd\theta \right]
\label{eq:refname21}
\end{eqnarray}

Since at $b=0$, $I_1=I_2$ and $\frac{\partial}{\partial b}(I_1)=-\frac{\partial}{\partial b}(I_2)$, therefore we have

\begin{eqnarray}
\left.\frac{\partial S}{\partial b}\right|_{b=0}
&=&\frac{2I_2\left.\frac{\partial}{\partial b}I_1\right|_{b=0}-2I_1\left.\frac{\partial}{\partial b}I_2\right|_{b=0}}{\left.(I_1+I_2)^2\right|_{b=0}}\nonumber \\
&=&\frac{2I_1\left.\frac{\partial}{\partial b}I_1\right|_{b=0}+2I_1\left.\frac{\partial}{\partial b}I_1\right|_{b=0}}{\left.(2I_1)^2\right|_{b=0}}\nonumber \\
&=&\frac{\left.\frac{\partial}{\partial b}I_1\right|_{b=0}}{\left. I_1\right|_{b=0}}
\label{eq:refname22}
\end{eqnarray}

Thus, equation \ref{eq:refname22} gives the sensitivity of the sensor. Now using equation \ref{eq:refname15a}, \ref{eq:refname16}, \ref{eq:refname21} and \ref{eq:refname22}, the sensitivity of the sensor can be calculated for different RMS amplitude of $Z_5$ mode ($a$) and also for different detector radius in Fourier plane coordinates ($\rho_p$) \cite{goodman2005introduction}.

\begin{figure}[H]
\centering
\includegraphics[scale=0.4]{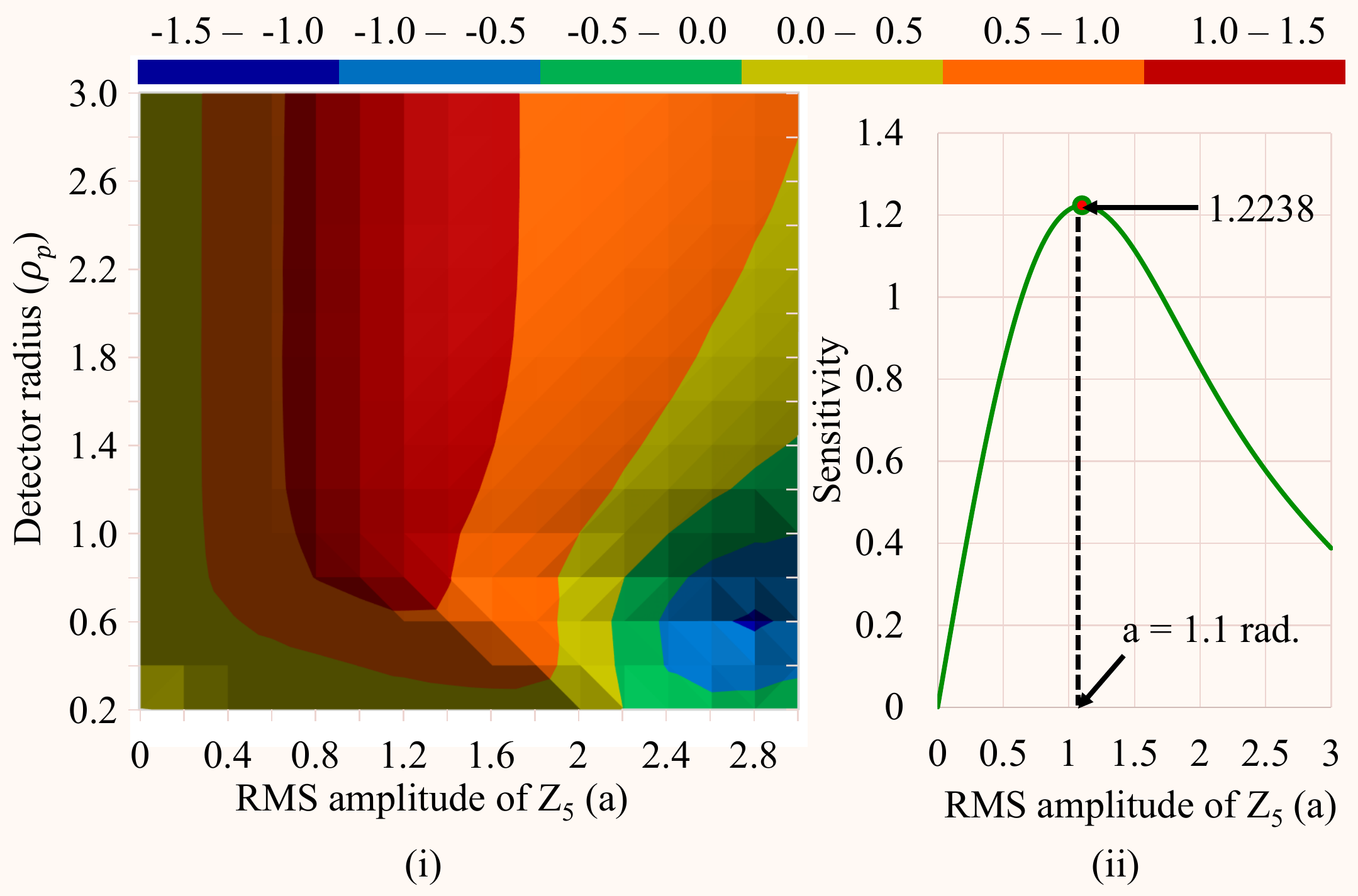}
\caption{Plot of sensitivity verses (i) RMS amplitude of astigmatism $45^o$ (i.e. $a$) in radian and detector radius in Fourier coordinate (i.e. $\rho_p$) and (ii) RMS amplitude of astigmatism $45^o$ (i.e. $a$) in radian.}
\label{fig3}
\end{figure}

Figure \ref{fig3}(i) shows the plot of the sensitivity of the sensor for different RMS amplitude of $Z_5$ mode ($a$) and detector radius in Fourier plane coordinates ($\rho_p$). In the figure $a$ is taken along the horizontal axis and $\rho_p$ is taken along the vertical axis, while the sensitivity is represented in different colors ranging from -1.5 to +1.5. The physical detector radius can be calculated from the Fourier plane coordinates ($\rho_p$) by using the expression $R=\rho_p \lambda f$, where $\lambda$ is the wavelength of the light beam and $f$ is the focal length of the focusing lens used. It is observed that for RMS amplitudes of $Z_5$ ranging from $a = 0.8$ – $1.4$ radian and $\rho_p > 0.8$ we get a comparatively good sensitivity (i.e. $> 1$). It is found that a sensitivity of $\approx 1.2$ is achieved at $\rho_p = 2.1$. Figure 3(ii) shows a plot of the sensitivity of the sensor verses RMS amplitudes of $Z_5$ ($a$) taking $\rho_p = 2.1$. It is observed that the maximum sensitivity of $1.2238$ is obtained at $a = 1.1$ radian.

\subsection{Cross-sensitivity of Astigmatic Defocus Sensor}

\begin{figure}[H]
\centering
\includegraphics[scale=0.5]{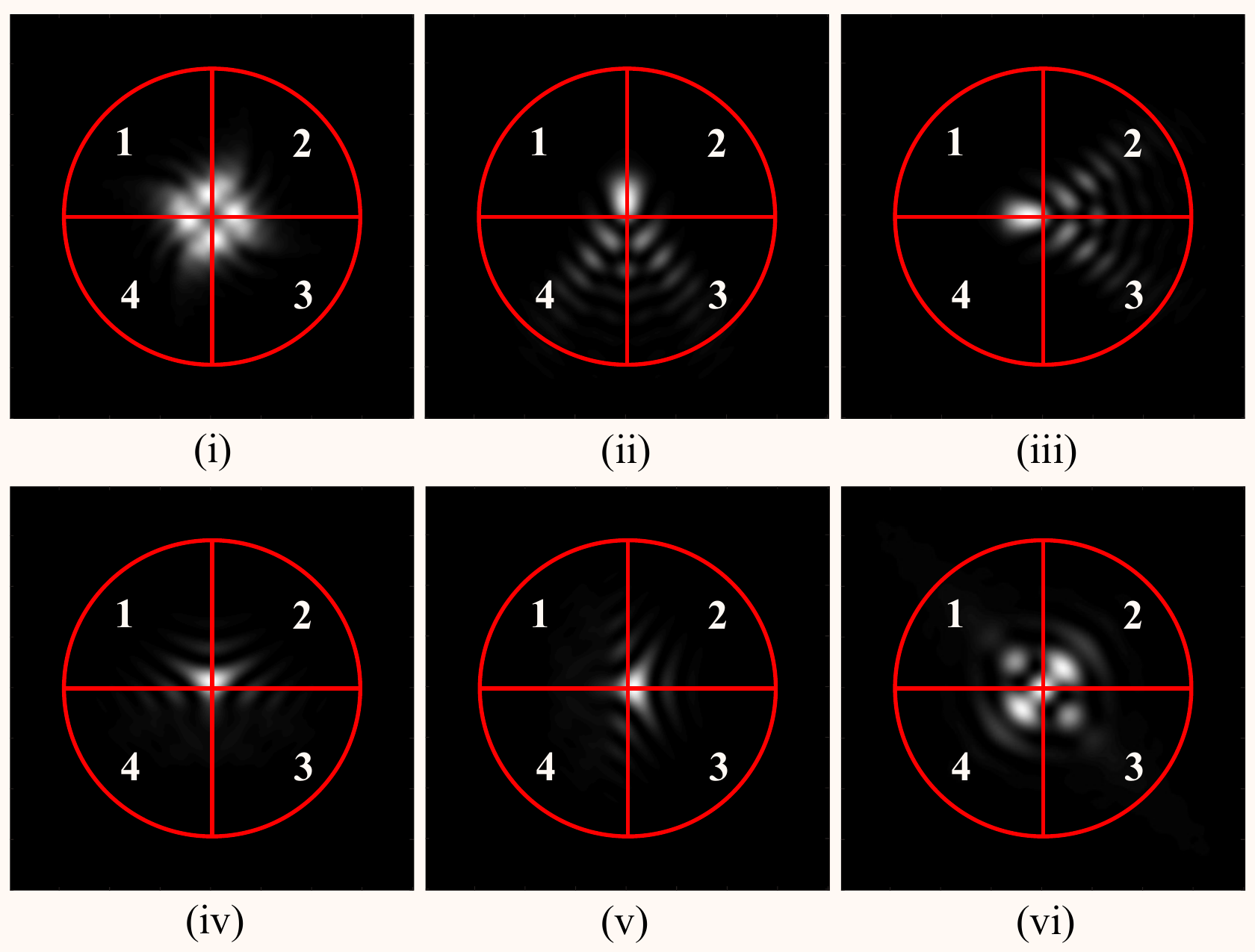}
\caption{Illustration of the intensity distribution at the detector plane with $a=1$ radian and $1$ radian RMS amplitude of (i) astigmatism $0^o$ ($Z_6$), (ii) y coma ($Z_7$), (iii) x coma ($Z_8$), (iv) y trefoil ($Z_9$), (v) x trefoil ($Z_{10}$) and (vi) spherical aberration ($Z_{11}$)}
\label{fig4}
\end{figure}

In order to test the effectiveness of the sensor, let us check whether the sensor shows any sensitivity towards other Zernike modes (aberration modes), i.e. other than the defocus mode. We termed it as cross-sensitivity. We tested the cross-sensitivity considering one of the six low order aberration modes to be present in the beam. The intensity distribution of the beam, comprising astigmatism $0^o$ ($Z_6$), y coma ($Z_7$), x coma ($Z_8$), y trefoil ($Z_9$), x trefoil ($Z_{10}$) and spherical aberration ($Z_{11}$) at the detector plane when focused by an astigmatic lens is shown in Fig. \ref{fig4} (i)$\rightarrow$ (vi).\\

The evaluation of the cross-sensitivity of each of the modes is discussed below -

\begin{itemize}
\item From Fig \ref{fig4}(i) it is seen that in case of astigmatism $0^o$ (Zernike mode $Z_6$) the intensities of sub-aperture 1, 2, 3, and 4 are equal, i.e. $I_1=I_2=I_3=I_4$ . Thus, from equation \ref{eq:refname17} the sensor output, $S=0$ and hence the sensitivity $\left. \frac{\partial S}{\partial b}\right|_{b=0}=0$.
\item From Fig \ref{fig4}(ii) it is seen that in case of y coma (Zernike mode $Z_7$) the intensity distribution of sub-aperture 1 $\&$ 2 and 3 $\&$ 4 are symmetric and thus $I_1=I_2$ and $I_3=I_4$. Therefore, from equation \ref{eq:refname17} the sensor output, $S=0$ and hence the sensitivity $\left. \frac{\partial S}{\partial b}\right|_{b=0}=0$.
\item From Fig \ref{fig4}(iii) it is seen that in case of x coma (Zernike mode $Z_8$) the intensity distribution of sub-aperture 1 $\&$ 4 and 2 $\&$ 3 are symmetric and thus $I_1=I_4$ and $I_2=I_3$. Therefore, from equation \ref{eq:refname17} the sensor output, $S=0$ and hence the sensitivity $\left. \frac{\partial S}{\partial b}\right|_{b=0}=0$.
\item From Fig \ref{fig4}(iv) it is seen that, similar to that of y coma, in case of y trefoil (Zernike mode $Z_9$) the intensity distribution of sub-aperture 1 $\&$ 2 and 3 $\&$ 4 are symmetric and thus $I_1=I_2$ and $I_3=I_4$. Therefore, from equation \ref{eq:refname17} the sensor output, $S=0$ and hence the sensitivity $\left. \frac{\partial S}{\partial b}\right|_{b=0}=0$.
\item From Fig \ref{fig4}(v) it is seen that, similar to that of x coma, in case of x trefoil (Zernike mode $Z_{10}$) the intensity distribution of sub-aperture 1 $\&$ 4 and 2 $\&$ 3 are symmetric and thus $I_1=I_4$ and $I_2=I_3$. Therefore, from equation \ref{eq:refname17} the sensor output, $S=0$ and hence the sensitivity $\left. \frac{\partial S}{\partial b}\right|_{b=0}=0$.
\item From Fig \ref{fig4}(vi) it is seen that in case of spherical aberration (Zernike mode $Z_{11}$) the intensity distribution of sub-aperture 1 $\&$ 3 and 2 $\&$ 4 are symmetric and thus $I_1=I_3$ and $I_2=I_4$. The situation here is similar to that of defocus (Zernike mode $Z_{4}$). Therefore, there will a non zero cross-sensitivity in case of spherical aberration. The cross-sensitivity can be calculated by replacing Zernike mode $Z_4$ by $Z_{11}$ in equation \ref{eq:refname16} and \ref{eq:refname21} and by using equation \ref{eq:refname15a} and \ref{eq:refname22}.
\end{itemize}

The cross-sensitivities due to the Zernike modes $Z_6$, $Z_7$, $Z_8$, $Z_9$, $Z_{10}$ and $Z_{11}$ are shown in Table \ref{tab:cross_sensitivity_astigmatic}.

\begin{table}[H]
\centering
\caption{\bf cross-sensitivity obtained in a astigmatic defocus sensor due to different Zernike modes}
\begin{tabular}{ccccccc}
\hline
Zernike modes & $Z_6$ & $Z_7$ & $Z_8$ & $Z_9$ & $Z_{10}$ & $Z_{11}$\\
\hline
Cross-sensitivity & 0 & 0 & 0 & 0 & 0 & -0.497\\
\hline
\end{tabular}
  \label{tab:cross_sensitivity_astigmatic}
\end{table}

It is observed from the Table \ref{tab:cross_sensitivity_astigmatic} that the sensitivities of all other modes ($Z_6$, $Z_7$, $Z_8$, $Z_9$ and $Z_{10}$ ) are zero, while in case of spherical aberration ($Z_{11}$) there is a maximum cross-sensitivity of -0.497, which is less than about $58\%$ from that of the sensitivity of the defocus aberration.

\section{Simulation}
Simulations are carried out in support of the above theoretical study. A plane wavefront of dimension $256 \times 256$ pixels is simulated. A fixed amount, say 0.001 radian RMS amplitude of Defocus aberration (i.e. Zernike mode $Z_4$) is added to the simulated plane wavefront. The beam with this wavefront now passes through an astigmatic lens which adds certain amount of astigmatism to the wavefront. Thus a certain amount $a$ of astigmatism $45^o$ (refer to table \ref{tab:zernike}) is further added to the wavefront. Now apart from adding astigmatism to the wavefront the astigmatic lens focuses the beam and the detector is placed at the focal plane of the lens. This can be achieved numerically through Fourier Transformation of the wavefront at the plane of the lens. The Fourier plane resembles the focal plane of the lens \cite{goodman2005introduction}. For better visibility a magnification factor of 20 along horizontal axis and another 20 along vertical axis, i.e. a total of 40, is incorporated in the Fourier plane.

\subsection{Sensitivity of Astigmatic Defocus Sensor}
The detector plane is divided as shown in Fig \ref{fig2} and equation \ref{eq:refname17} is used to obtain the sensor output. Now, since the sensitivity of a sensor is a measure of the effectiveness of the sensor while measuring small aberration and we have already incorporated a very small aberration (i.e. 0.001 radian RMS amplitude of $Z_4$) into the wavefront, the sensor output divided by 0.001 (i.e. $S/0.001$) gives the sensitivity of the sensor. The sensitivity of the sensor is then calculated for different value of $a$ (i.e. RMS amplitude of astigmatism $45^o$) and plotted with sensitivity along vertical axis and $a$ along horizontal axis.\\

\begin{figure}[H]
\centering
\includegraphics[scale=0.42]{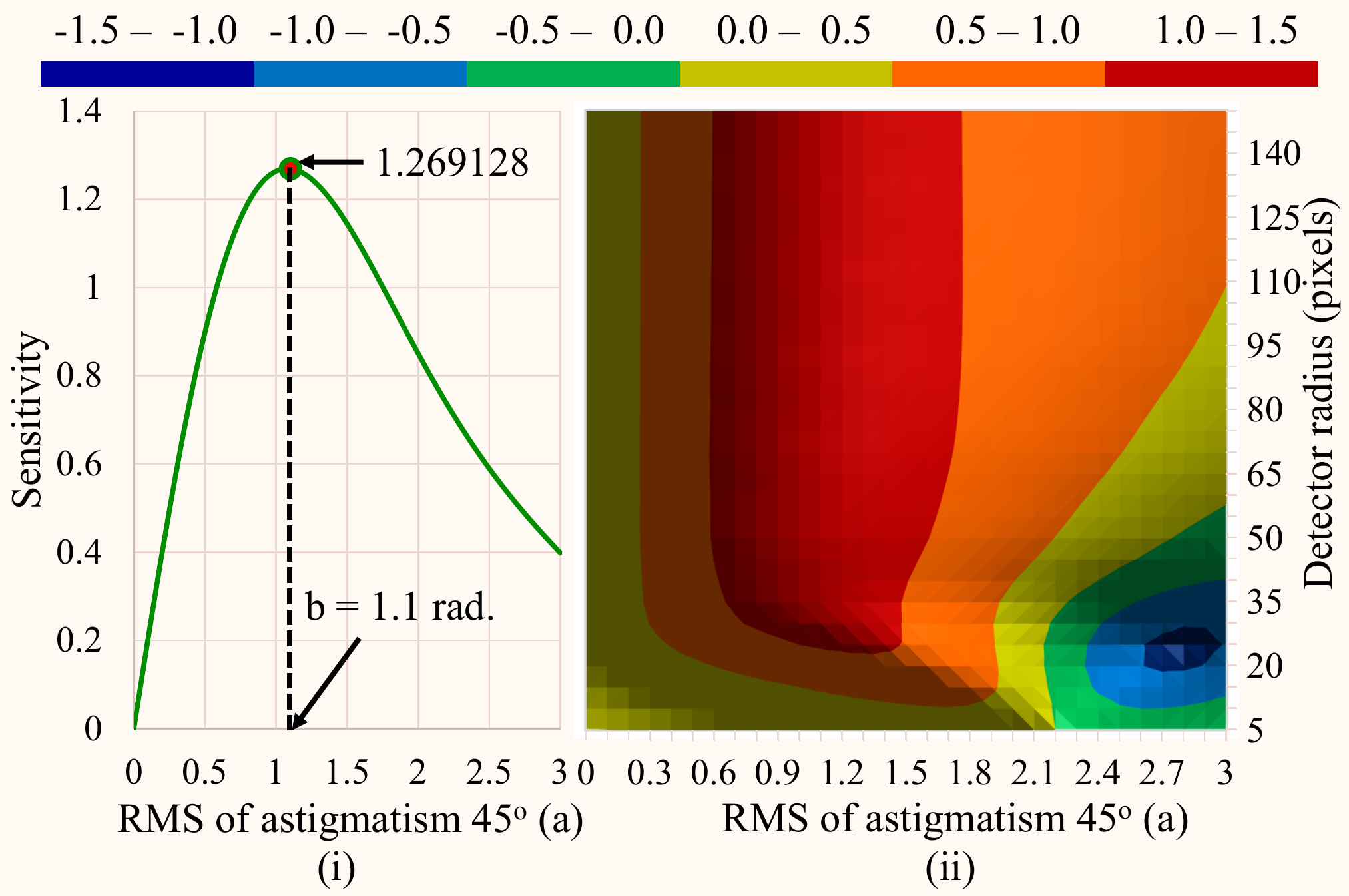}
\caption{Simulation plot of sensitivity verses (i) RMS amplitude of astigmatism $45^o$ (i.e. $a$) in radian and (ii) RMS amplitude of astigmatism $45^o$ (i.e. $a$) in radian and detector radius in number of pixels.}
\label{fig7}
\end{figure}

\begin{figure}[H]
\centering
\includegraphics[scale=0.42]{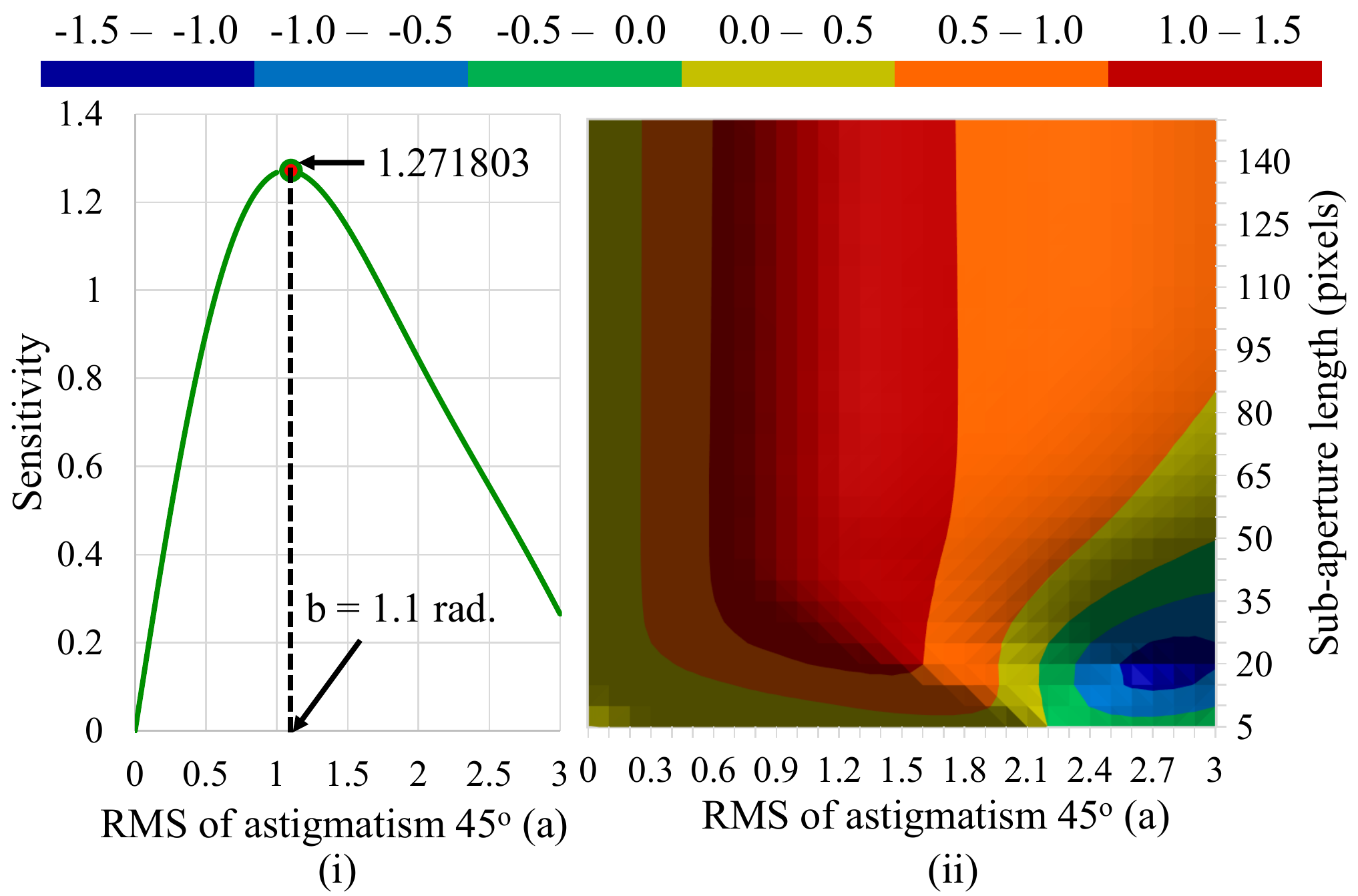}
\caption{Simulation plot of sensitivity verses (i) RMS amplitude of astigmatism $45^o$ (i.e. $a$) in radian and (ii) RMS amplitude of astigmatism $45^o$ (i.e. $a$) in radian and length of a sub-aperture of a square detector in number of pixels.}
\label{fig8}
\end{figure}

The plot of sensitivity verses RMS amplitude of astigmatism $45^o$ (i.e. $a$) in radian is shown in Fig \ref{fig7}(i) (in case of a detector with circular aperture) and in Fig \ref{fig8}(i) (in case of a detector with square aperture). Figure \ref{fig7}(ii) shows the plot of sensitivity verses RMS amplitude of astigmatism $45^o$ (i.e. $a$) in radian and detector radius of a circular aperture in number of pixels. The sensitivity is represented in different colors ranging from -1.5 to +1.5. The plot is obtained by varying the value of $a$ as well as the number of pixels defining the radius of the detector aperture.\\

It is observed that for RMS amplitudes of $Z_5$ ranging from $a = 0.8$ – $1.4$ radian and radius $ > 30$ pixels we get a comparatively good sensitivity (i.e. $> 1$). It is found that a sensitivity of $\approx 1.3$ is achieved at a detector radius of $85$ pixels. Therefore in the plot shown in Fig. \ref{fig7}(i), i.e. the plot of the sensitivity of the sensor verses RMS amplitudes of $Z_5$ ($a$), the radius of the detector aperture is taken as $85$ pixels. It is observed that the maximum sensitivity of $1.269128$ is obtained at $a = 1.1$ radian.

Similarly, Fig. \ref{fig8}(ii) shows the plot of sensitivity verses RMS amplitude of astigmatism $45^o$ (i.e. $a$) in radian and number of pixels defining the length of a sub-aperture of a detector with square aperture. The sensitivity is represented in different colors ranging from -1.5 to +1.5. The plot is obtained by varying the value of $a$ as well as the number of pixels defining the length of a sub-aperture of the detector.

In this case it is observed that for RMS amplitudes of $Z_5$ ranging from $a = 0.6$ – $1.7$ radian and length of a sub-aperture $ > 50$ pixels we get a comparatively good sensitivity (i.e. $> 1$). It is found that a sensitivity of $\approx 1.3$ is achieved at a sub-aperture length of $60$ pixels. Figure \ref{fig8}(i) shows a plot of the sensitivity of the sensor verses RMS amplitudes of $Z_5$ ($a$) taking the length of a sub-aperture of the detector $ = 60$ pixels. It is observed that the maximum sensitivity of $1.271803$ is obtained at $a = 1.1$ radian.\\

\begin{figure}[H]
\centering
\includegraphics[scale=0.44]{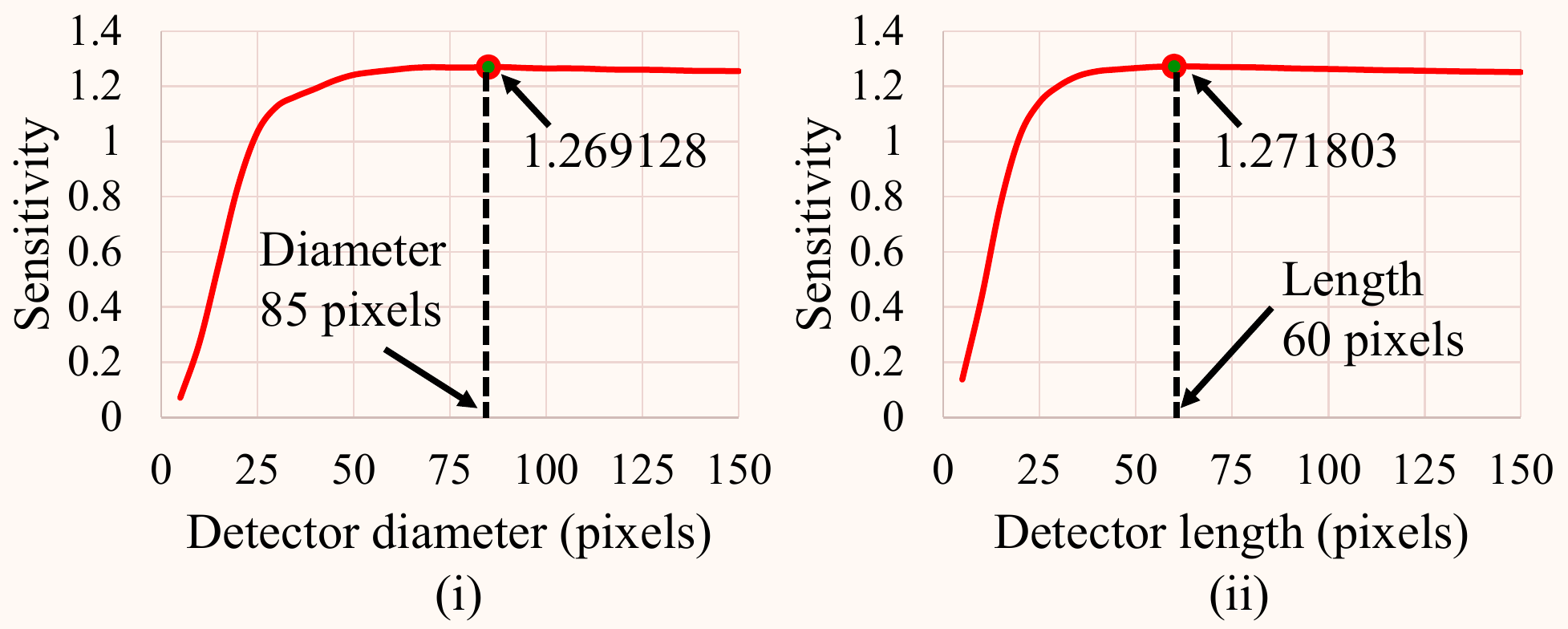}
\caption{Simulation plot of sensitivity verses (i) detector radius of a circular aperture and (ii) detector length of a square sub-aperture in number of pixels when RMS amplitude of the bias aberration, i.e. a = 1.1 radian.}
\label{fig9}
\end{figure}

Figure \ref{fig9} shows the simulation plot of sensitivity verses detector dimension in number of pixels when RMS amplitude of the bias aberration, i.e. a is 1.1 radian. Here, Fig. \ref{fig9}(i) is the plot of sensitivities obtained for a circular aperture of different radius while Fig. \ref{fig9}(ii) is the plot of sensitivities obtained for a square aperture of different sub-aperture lengths. It is seen that for a bias aberration of 1.1 radian a sensitivity of approximately 1.2 and higher can be achieved using a detector of circular aperture of radius equal to or greater than 40 pixels while the same can be achieved using a detector of square aperture of sub-aperture length equal to or greater than 30 pixels. It is also observed that, with increase in the dimension of the detector there is a very slight (almost negligible) and gradual decrease in the sensitivity after obtaining a peak value at a radius of 85 pixels for a circular aperture and a sub-aperture length of 60 pixels for a square aperture.

\subsection{Cross-sensitivity of Astigmatic Defocus Sensor}
The cross-sensitivity of the Astigmatic Defocus Sensor is checked through simulation by introducing different aberration modes into the incident wavefront. The simulation for cross-sensitivity is carried out considering $a=1.1$ radian and the radius of circular aperture equal to 60 pixels. Table \ref{tab:cross_sensitivity_astigmatic} shows the cross-sensitivity (C.S.) observed in a astigmatic defocus sensor due to the presence of different aberration modes (Zernike modes) in the test wavefront. It is observed that apart from $Z_{11}$ mode all other modes shows a very low cross-sensitivity ($< 0.007$). The $Z_{11}$ mode due to its symmetric resemblance with the $Z_{4}$ (i.e. defocus) mode shows a higher cross-sensitivity. The table so obtained through simulation is very much in agreement to that of the theoretical prediction.

\begin{table}[H]
\centering
\caption{\bf Cross-sensitivity, obtained through simulation, in a Astigmatic Defocus Sensor due to the presence of different Zernike modes in the test beam}
\begin{tabular}{ccccccc}
\hline
$Z_k$ & $Z_6$ & $Z_7$ & $Z_8$ & $Z_9$ & $Z_{10}$ & $Z_{11}$\\
\hline
C.S. & 0.0061 & 0.0063 & 0.0063 & 0.0061 & 0.0061 & -0.3499\\
\hline
\end{tabular}
  \label{tab:cross_sensitivity_astigmatic_sim}
\end{table}

\begin{figure}[H]
\centering
\includegraphics[scale=0.5]{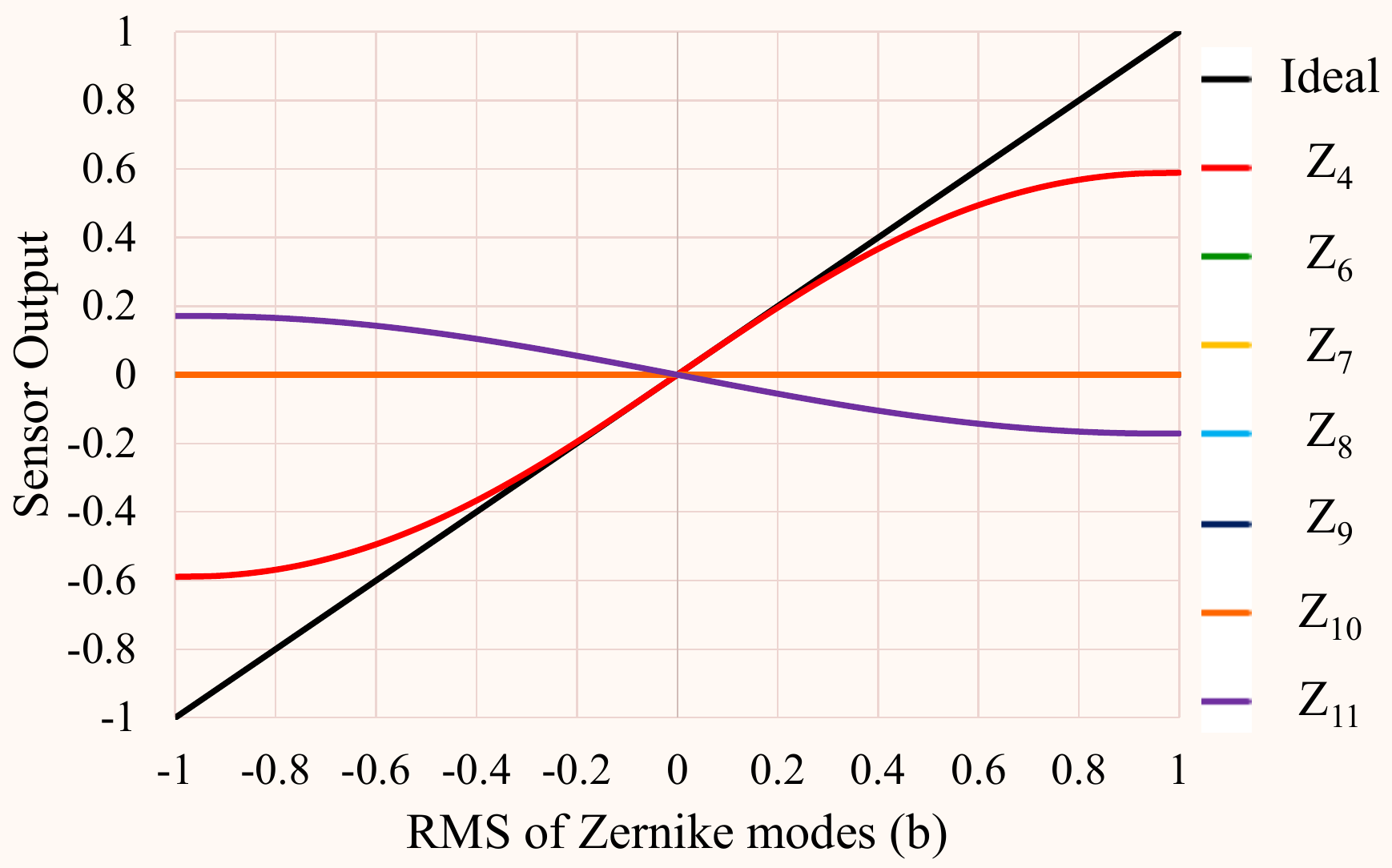}
\caption{Simulation plot of sensor output verses RMS amplitude (i.e. $b$ in radian) of Zernike modes $Z_k$ for $k = 4 \rightarrow 11$.}
\label{fig10}
\end{figure}

Figure \ref{fig10} shows the response of the Astigmatic Defocus Sensor towards the presence of different Zernike modes. These are the plots of the sensor output verses RMS amplitude of the Zernike modes $Z_4$ to $Z_{11}$ taking bias (a) as 1.1 radian and radius as 60 pixels, i.e. for good sensitivity. From the figure we can see that the astigmatic defocus sensor shows a good response to the defocus mode ($Z_4$) as the curve coincides with the ideal curve to some extent. However, the sensor shows no response to Zernike modes $Z_6$, $Z_7$, $Z_8$, $Z_9$ and $Z_{10}$ as the curve is a near straight line at sensor output '0'. Further, it shows a very low but distinct response to Zernike modes $Z_{11}$ but in an opposite sense to that of the defocus mode ($Z_4$). This is in agreement with the above table as well as the theoretical prediction.

\subsection{Linear response of Astigmatic Defocus Sensor}
\begin{figure}[H]
\centering
\includegraphics[scale=0.5]{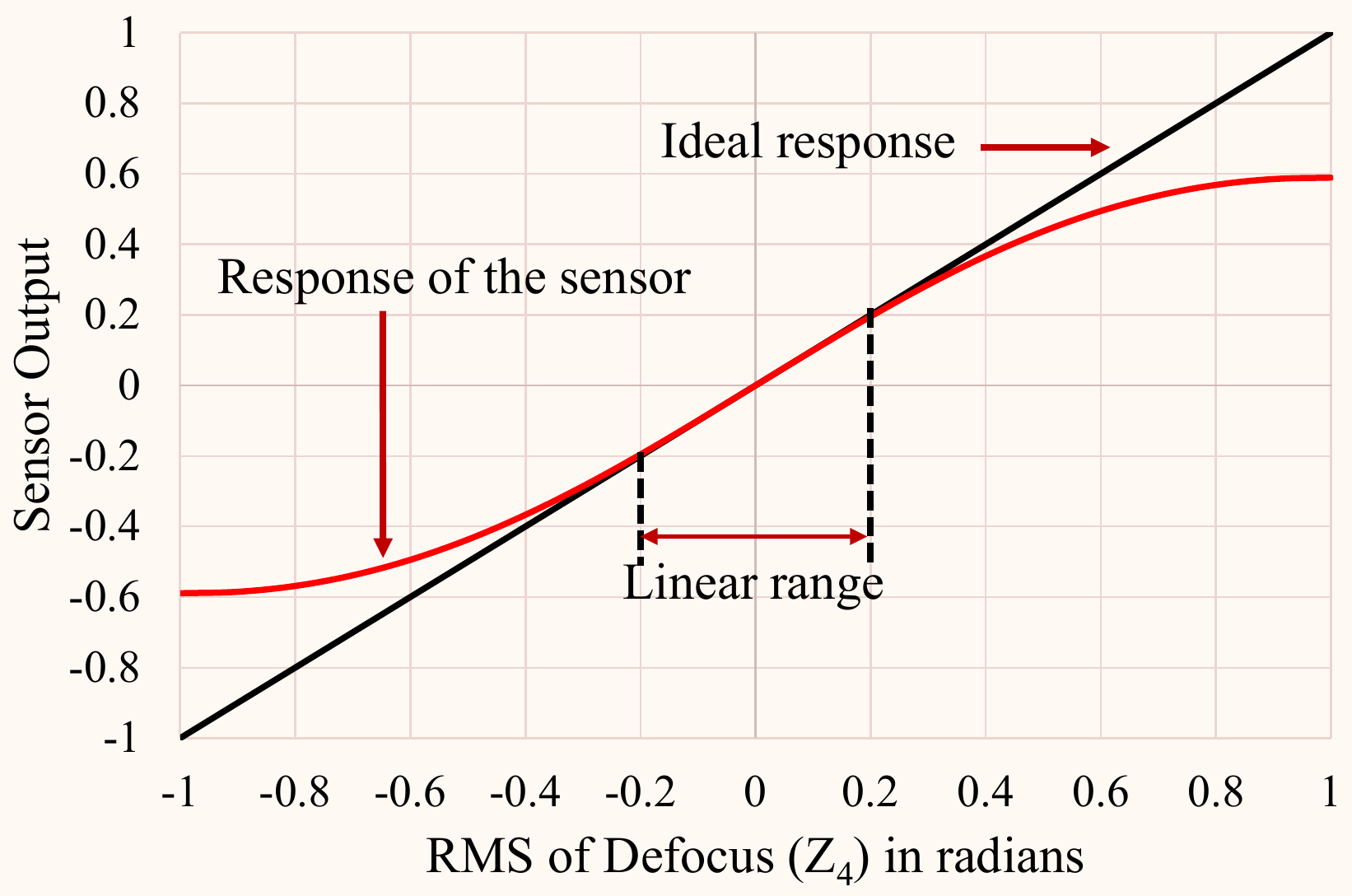}
\caption{Simulation plot of sensor output verses RMS amplitude (i.e. $b$ in radian) of defocus aberration mode (Zernike mode $Z_4$) present in the input beam.}
\label{fig11}
\end{figure}

Liner response is a measure of how linear is the change in sensor output to a linear change in the input to the sensor. Figure \ref{fig11} shows the simulation plot of the response curve of the sensor and the ideal response. These are basically the plots of the sensor output verses RMS amplitude (i.e. $b$ in radian) of defocus aberration mode (Zernike mode $Z_4$) present in the input beam.\\

The sensor can be considered as showing a linear response, if its response curve coincides with the ideal response curve. It is observed that the response curve of the sensor coincides with the ideal response curve for a limited range on the either side of the zero RMS amplitude of $Z_4$. The response curve of the sensor shown in Fig \ref{fig11} coincides with the ideal curve from RMS amplitude of -0.2 radian to +0.2 radian approximately, i.e. linear for a range of 0.2 radian of input aberration mode.\\

\begin{figure}[H]
\centering
\includegraphics[scale=0.42]{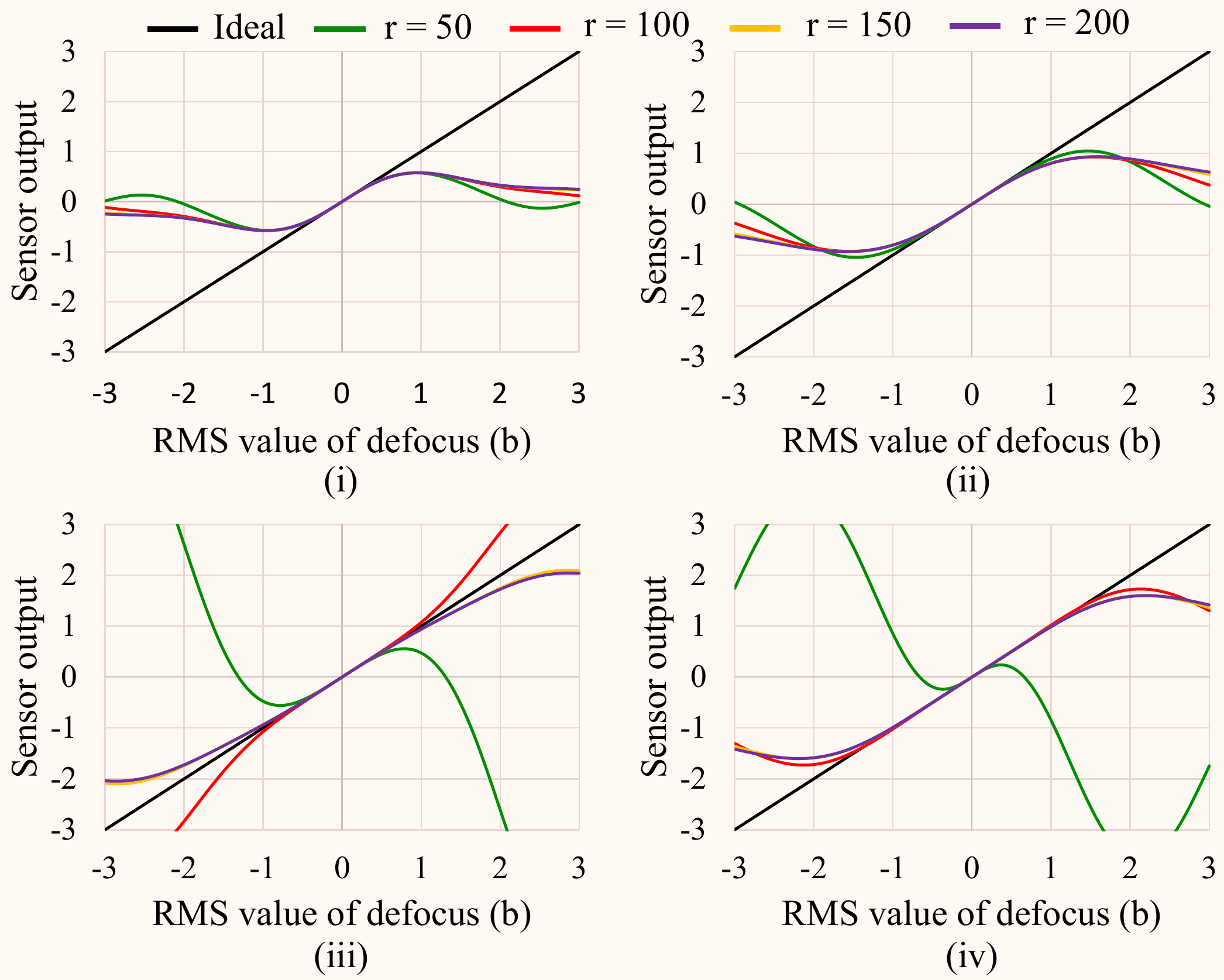}
\caption{Simulation plot of sensor output verses RMS amplitude (i.e. $b$ in radian) of defocus mode (Zernike mode $Z_4$) for RMS amplitude of the bias mode (Zernike mode $Z_5$) (i) $a = 1$ radian, (ii) $a = 2$ radians, (iii) $a = 3$ radians and (iv) $a = 4$ radians with a circular detector of radius (r) of 50, 100, 150 and 200 pixels}
\label{fig12}
\end{figure}

Figure \ref{fig12} shows the response curves of the sensor with a circular aperture of different radius and with different RMS amplitude of the bias aberration mode (i.e. Zernike mode $Z_5$). Figure \ref{fig12}(i $\rightarrow$ iv) are plots of the response curve with RMS amplitude of bias $a = 1, 2, 3$ and $4$ radians respectively. The response curve with aperture of different radius ($r$) are represented by different colours, i.e. black are the ideal curves, green are curves for $r = 50$ pixels, red are curves for $r = 100$ pixels, yellow are curves for $r = 150$ pixels and violet are curves for $r = 200$ pixels.\\

\begin{figure}[H]
\centering
\includegraphics[scale=0.42]{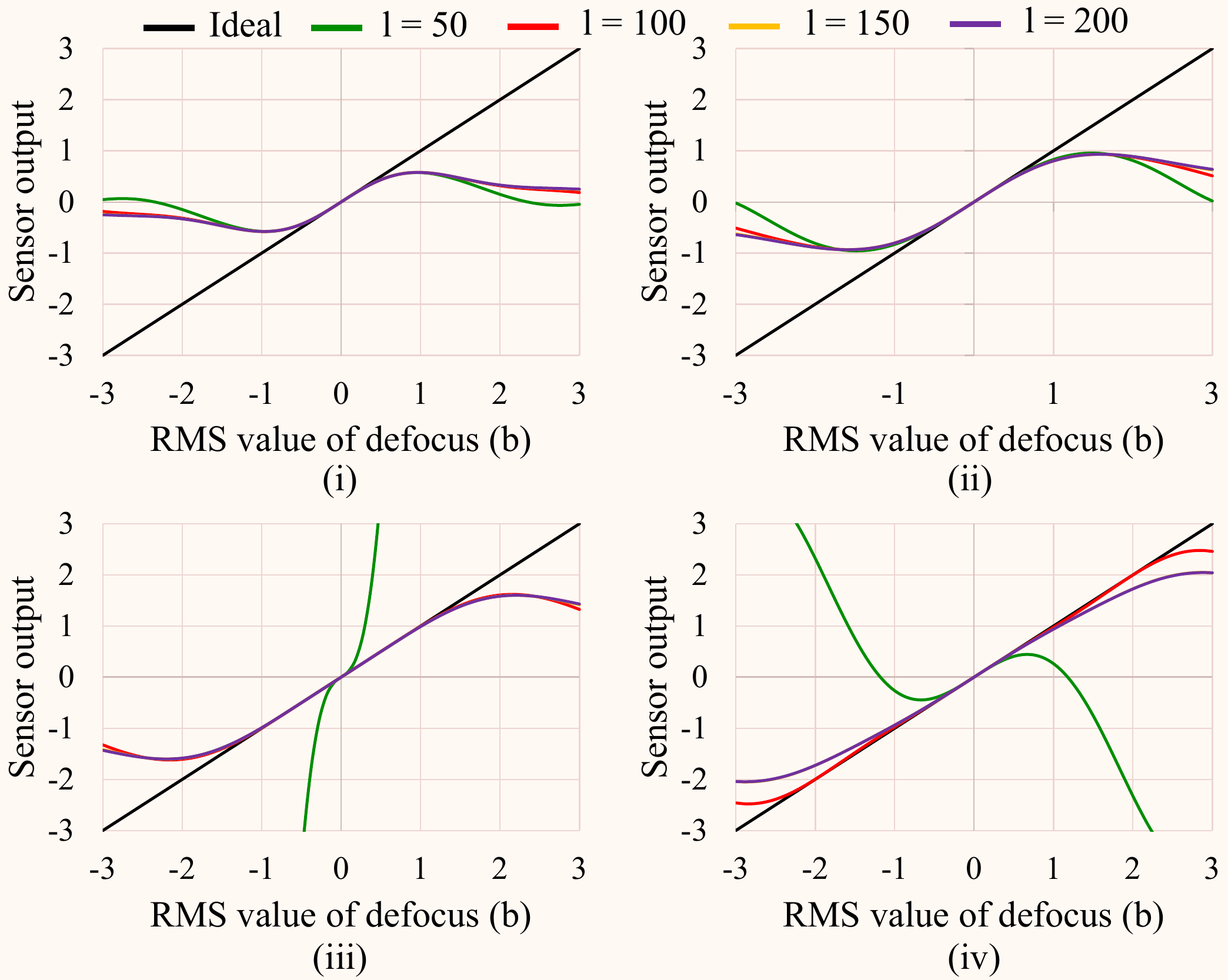}
\caption{Simulation plot of sensor output verses RMS amplitude (i.e. $b$ in radian) of defocus mode (Zernike mode $Z_4$) for RMS amplitude of the bias mode (Zernike mode $Z_5$) (i) $a = 1$ radian, (ii) $a = 2$ radians, (iii) $a = 3$ radians and (iv) $a = 4$ radians with a square detector sub-aperture length (l) of 50, 100, 150 and 200 pixels}
\label{fig13}
\end{figure}

Similarly, Fig. \ref{fig13} shows the response curves of the sensor with a square aperture of different sub-aperture lengths and with different RMS amplitude of the bias aberration mode (i.e. Zernike mode $Z_5$). Figure \ref{fig13}(i $\rightarrow$ iv) are plots of the response curve with RMS amplitude of bias $a = 1$, $2$, $3$ and $4$ radians respectively. The response curve with sub-aperture of different lengths ($l$) are represented by different colours, i.e. black, green, red, yellow and violet are the ideal curves, curves for $l = 50$ pixels, curves for $l = 100$ pixels, curves for $l = 150$ pixels and curves for $l = 200$ pixels respectively.

\begin{figure}[H]
\centering
\includegraphics[scale=0.5]{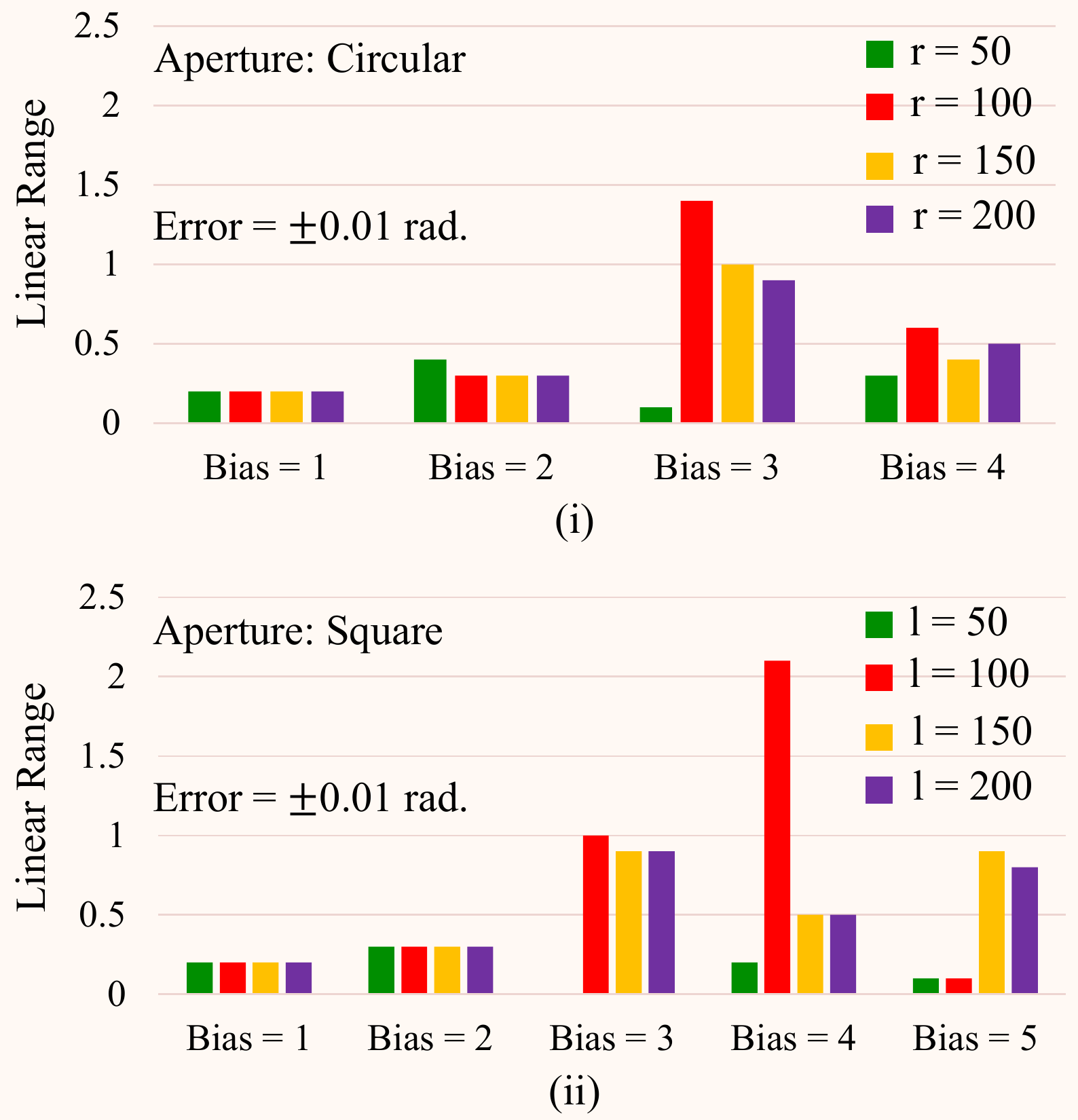}
\caption{Bar diagram of the linear range of the response curves shown in Fig. \ref{fig12} and Fig. \ref{fig13} in case of (i) a circular aperture and (ii) a square aperture.}
\label{fig14}
\end{figure}

Figure \ref{fig14}(i) shows the bar diagram of the linear range of the response curves shown in Fig. \ref{fig12} i.e. for a circular aperture, at an error of $\pm 0.01$ radian. The different set of bars with different colours are for bias values $a = 1$, $2$, $3$ and $4$ radians respectively while bars of different colour represents the linear range for circular aperture of different radius. The colour codes are same as that of Fig. \ref{fig12}. Similarly, Fig. \ref{fig14}(ii) shows the bar diagram of the linear range of the response curves shown in Fig. \ref{fig13} i.e. for a square aperture, at an error of $\pm 0.01$ radian. The different set of bars with different colours are for bias values $a = 1$, $2$, $3$ and $4$ radians respectively while bars of different colour represents the linear range for square aperture of different sub-aperture lengths. The colour codes are same as that of Fig. \ref{fig13}.\\

\begin{figure}[H]
\centering
\includegraphics[scale=0.5]{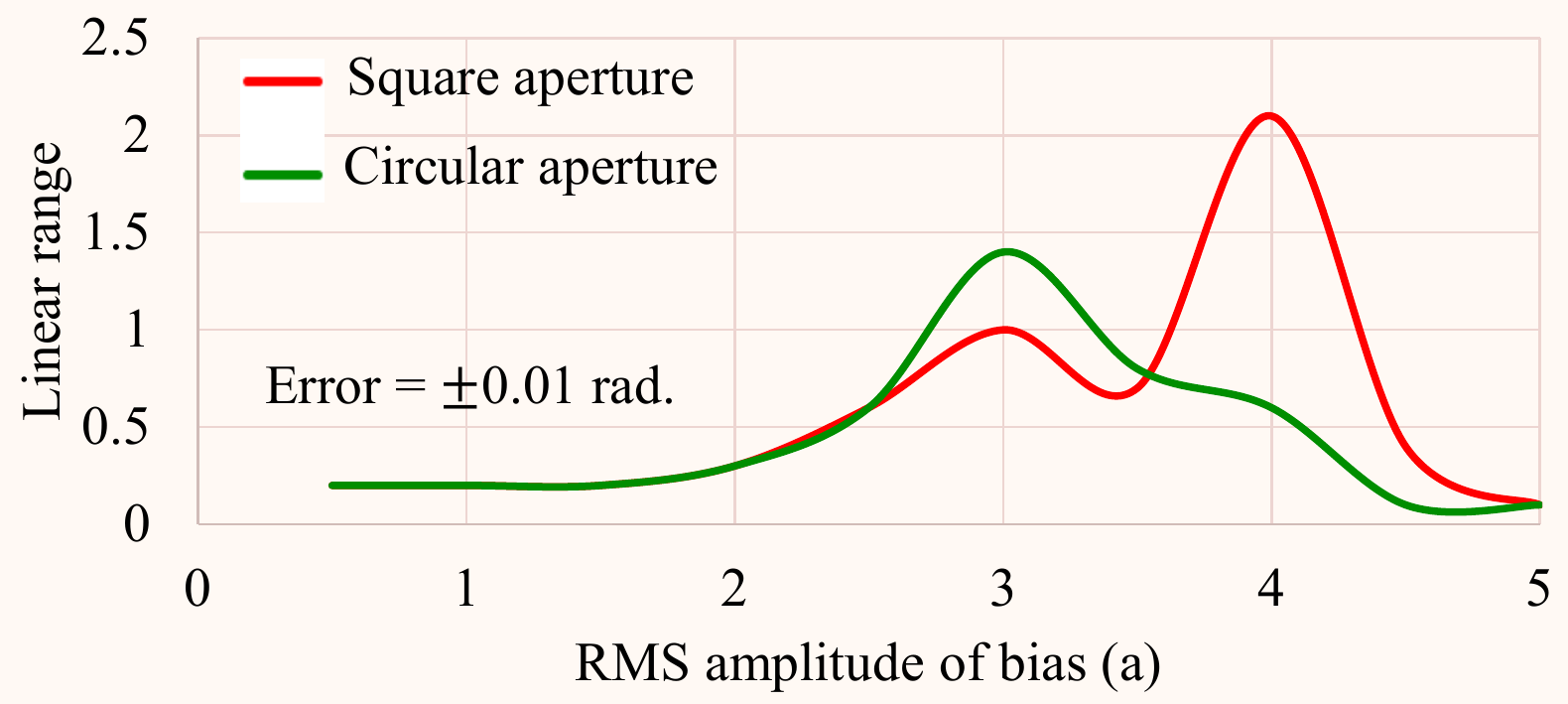}
\caption{Plot of linear range verses RMS amplitude of the bias aberration (b) for an error of $\pm 0.01$ radians.}
\label{fig15}
\end{figure}

From Fig. \ref{fig12}, \ref{fig13} and \ref{fig14} it is found that a maximum linear range of 1.4 is achieved for a detector radius of 100 pixels in case of circular aperture while a maximum linear range of 2.1 is achieved for a sub-aperture length of 100 pixels in case of square aperture. Further, it is also seen that the linear range irrespective of the shape of the detector aperture increase gradually with increase in the amplitude of the bias aberration, reaches a peak and then falls back (clearly shown in Fig. \ref{fig15}). Figure \ref{fig15} shows the plot of linear range verses RMS amplitude of the bias aberration (b) for an error of $\pm 0.01$ radians for both circular (green curve) and square (red curve) aperture. It is observed that a maximum linear range (peak of the curves) is obtained at a bias amplitude of 3 radian in case of a circular aperture and at a bias amplitude of 4 radian in case of a square aperture.

\section{Performance Optimization of the Astigmatic Defocus Sensor}
As seen in the above discussions, the performance of the Astigmatic Defocus Sensor can be optimized both in terms of its sensitivity as well as its linear response. Let us consider both the cases one by one.

\subsection{Optimization for higher Sensitivity}
It is observed that in order to achieve a good sensitivity (i.e. $> 1$), the RMS amplitude of $Z_5$ is to be kept in between $a = 0.8$ – $1.4$ radian and radius $r > 60$ pixels in case of a circular aperture, while the same can be achieved for an RMS amplitude of $Z_5$ in between $a = 0.6$ – $1.7$ radian and length of a side $l > 50$ pixels in case of a square aperture. It is also observed that, irrespective of the shape of the detector aperture, there is a very small change in the sensitivities with increase in the dimension of the detector. However, a maximum sensitivity is observed at a radius of 85 pixels for a circular aperture and a length of 60 pixels for a square aperture with amplitude of the bias aberration $b = 1.1$ radian in both the case. Further a sensor with square shaped aperture shows a slightly higher sensitivity (sensitivity = 1.271803) than that of a sensor with circular shaped aperture (sensitivity = 1.269128)\\

Thus a sensor with comparatively good sensitivity can be designed by keeping the detector dimension larger that 50 pixels and the RMS amplitude of the bias aberration (i.e. astigmatism) at around 1.1 radian irrespective of the shape of the detector aperture.

\subsection{Optimization for higher linearity}
It is observed that a maximum linear range is achieved for a detector dimension of 200 pixels irrespective of the shape of the detector aperture.  Further, a maximum linear range of 1.4 is achieved for an amplitude of 3 radian of the bias aberration in case of circular aperture while a maximum linear range of 2.1 is achieved for an amplitude of 4 radian of the bias aberration in case of square aperture. Thus a sensor with a square shaped detector aperture gives a better linear response compared to that of a sensor with a circular shaped detector aperture.\\

Thus a sensor with comparatively good linear response can be designed by keeping the detector dimension at around 200 pixels and the RMS amplitude of the bias aberration (i.e. astigmatism) at around 3 radian in case of a circular shaped detector aperture and at around 4 radian in case of a square shaped detector aperture. The linear response can be further improved by choosing a square shaped detector aperture rather than a circular one.

\section{Implementation of Astigmatic Defocus Sensor using Computer Generated Hologram}
The Astigmatic Defocus Sensor can be easily implemented using computer generated hologram (CGH). A CGH is the record of a numerically computed interference pattern based on the known information of the phase profile of the object beam (the beam assumed to be reflected from an object) and the reference beam (normally a beam with a plane wavefront)\cite{konwar2019performance}. When a beam with a phase profile same as that of the reference beam pass through the CGH and focused by a lens, the beam is diffracted giving rise to diffraction orders, such that a zero order at the center and other higher orders ($\pm 1, \pm 3, \pm 5 ...$) on the either side of the zero order. Among these diffraction orders, the $+1$ order carries the phase profile of the object beam. A tilt function is used to locate the diffraction orders at spatially different positions\cite{boruah2009dynamic}. This property of the CGH can be used to manipulate the phase profile of a light beam.\\

\begin{figure}[H]
\centering
\includegraphics[scale=0.5]{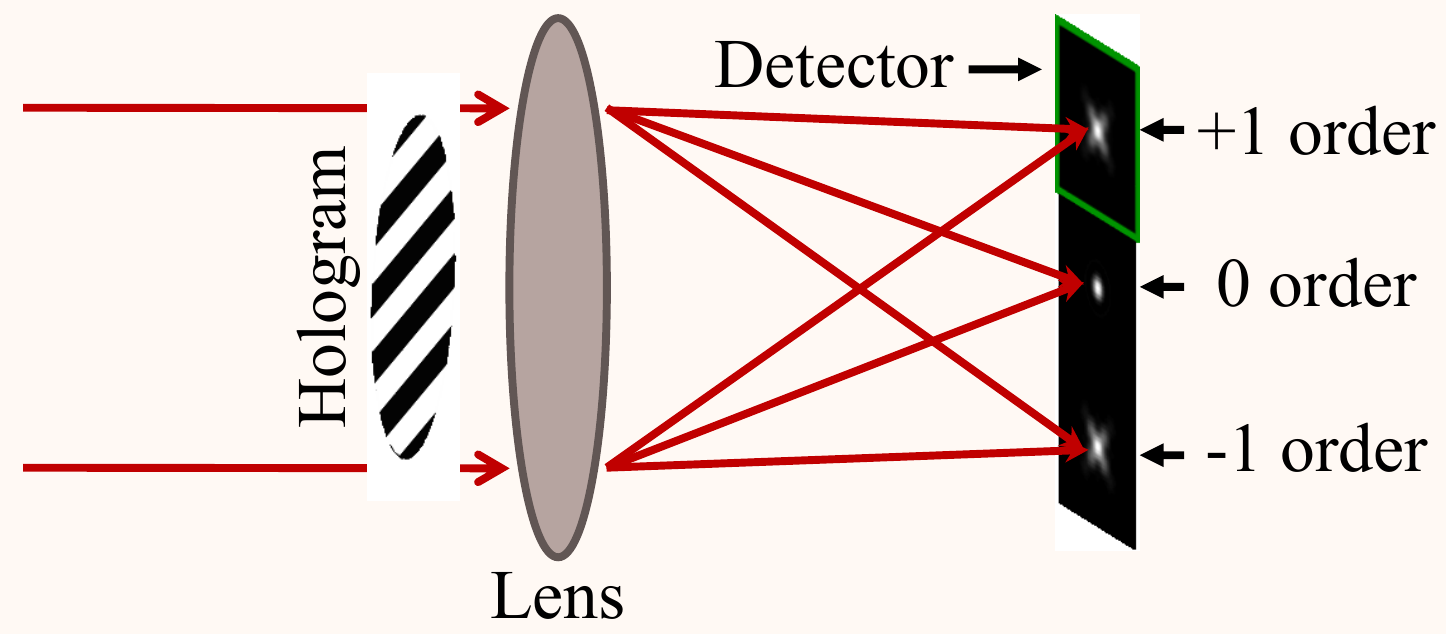}
\caption{Illustration of the basic principle of a Holographic Astigmatic Defocus Sensor.}
\label{fig16}
\end{figure}

\begin{figure}[H]
\centering
\includegraphics[scale=0.5]{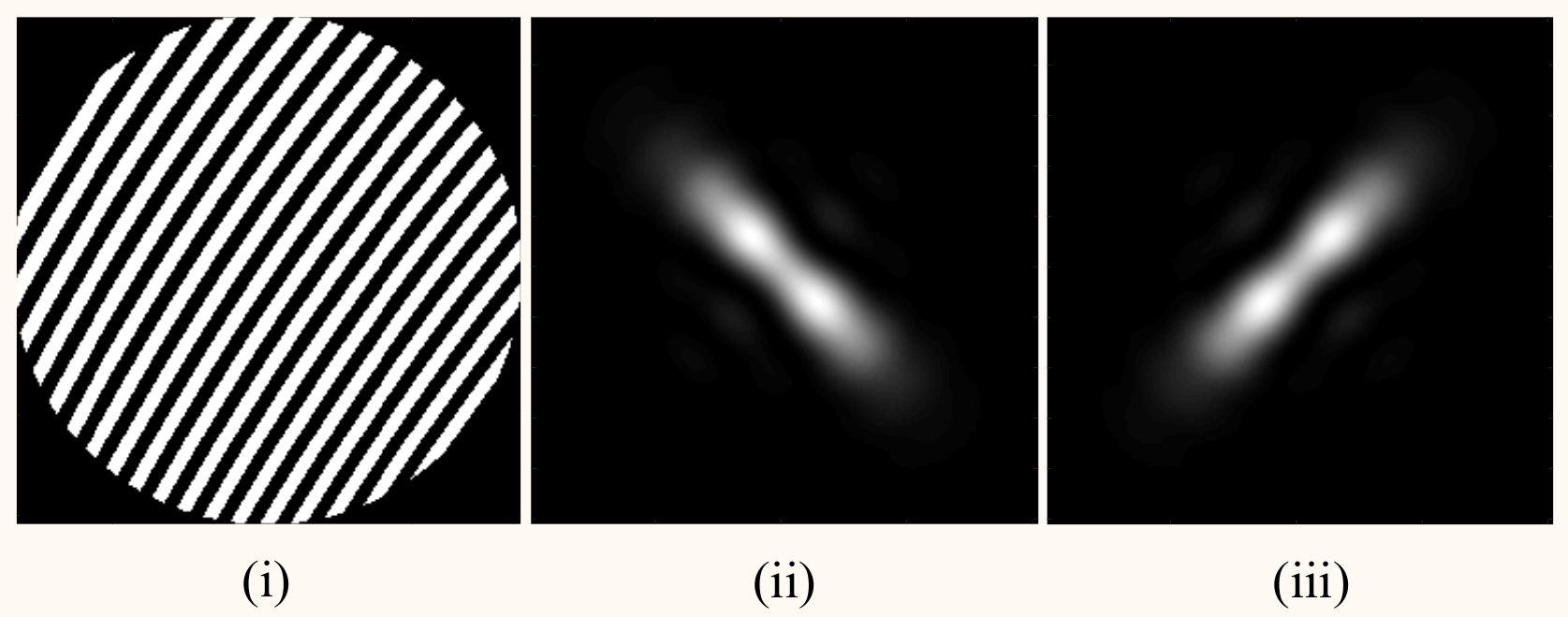}
\caption{Illustration of (i) a Computer Generated Hologram and intensity distribution of the +1 order for a beam with (ii) 1 radian and (iii) -1 radian of Defocus aberration.}
\label{fig17}
\end{figure}

The basic principle of implementing the Astigmatic Defocus Sensor using CHG is shown in Fig \ref{fig16}. Since the sensor is designed using hologram, it can also be called a Holographic Astigmatic Defocus Sensor. Here the astigmatic lens is replaced by a CGH and a converging lens immediately after it as shown. If the CGH is designed with Zernike mode $Z_5$ as the object beam phase profile and a plane wavefront as reference beam phase profile, we get an astigmatic intensity profile at the $+1$ diffracted order. A detector (bordered green in Fig \ref{fig16}) is used to capture the $+1$ diffracted order. If +1 radian RMS amplitude of defocus aberration is present in the beam incident on the CGH, intensity distribution of the +1 order at the detector plane will appear as shown in Fig \ref{fig17}(ii). Similarly, if -1 RMS amplitude of defocus aberration is present in the beam incident on the CGH, intensity distribution of the +1 order at the detector plane will appear as shown in Fig \ref{fig17}(iii) respectively. Figure \ref{fig17}(i) shows a representative CGH placed at the hologram plane. The process of obtaining the amount of defocus present in the incident beam is similar to that of the conventional Astigmatic Defocus Sensor discussed above.

\begin{figure}[H]
\centering
\includegraphics[scale=0.5]{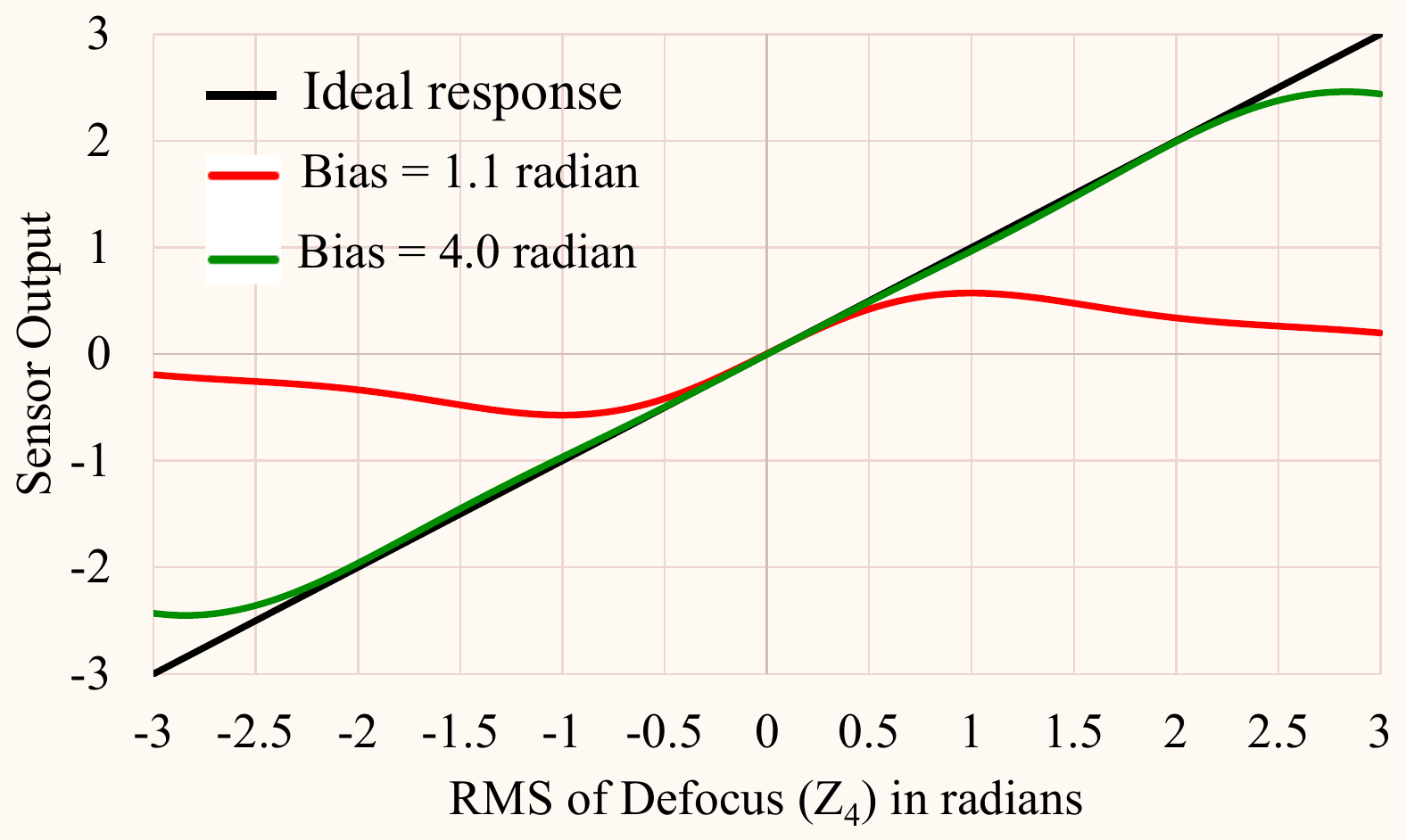}
\caption{Response curves of a Holographic Astigmatic Defocus Sensor for bias $a = 1.1$ radian (red), $a = 4.0$ radian (green) and the ideal response (black).}
\label{fig18}
\end{figure}

Figure \ref{fig18} shows the simulated response curve of the Holographic Astigmatic Defocus Sensor. The response curves are obtained for a detector of square aperture of sub aperture length $l = 100$ pixels. The black curve represents the ideal response, the red curve is the response curve of the sensor for bias $a = 1.1$ radian (i.e. for maximum sensitivity) and the green curve is the response curve of the sensor for bias $a = 4.0$ radian (i.e. for maximum linearity). It is clear form the figure that the response curves are in agreement with the response of the conventional sensor, i.e. the red curves of Fig. \ref{fig13}(i) and \ref{fig13}(iv).

\section{Conclusion}
In this paper we put forward a theoretical explanation of the principle of an Astigmatic Defocus Sensor using the concept of Fourier optics. The Zernike polynomials are used to represent the aberration modes. An expression for sensitivity of the sensor is also established. The effect of the presence of other aberration modes, i.e. other than defocus mode, is explored by obtaining the cross-sensitivity due to the other modes. A significant affect on the sensor output is seen due to the presence of the spherical aberration in the beam, while the presence of other modes is insignificant. A simulation study on the sensitivity, cross-sensitivity and linear response of the sensor is also carried out. The plots obtained from the theoretical expressions are in agreement with the simulation plots.\\

Further, the criteria for optimizing the sensor in terms of sensitivity and linear response is also discussed. It is found that higher sensitivities can be achieved by keeping the amplitude of the bias aberration at 1.1 radian while higher linear response can be achieved by keeping the amplitude of the bias aberration at 3 radian for a circular aperture and 4 radian for a square aperture. Further, a sensor designed for higher sensitivity will give low linear response and vice versa. It is also observed that there is a very little or no effect of the shape of the detector on the sensitivities, however, a comparatively higher linear response can be achieved by using a detector of square aperture than that of a circular aperture.\\

Later, we discussed how the sensor can be implemented using computer generated holograms. The response of such Holographic Astigmatic Defocus Sensor is in agreement with the conventional one.

% Bibliography
\bibliography{references}

\end{multicols}

\pagebreak
\begin{center}
{\large \bf Annexure 1}
\end{center}

\begin{multicols}{2}
Now let us see how we can obtain the equation
\begin{eqnarray}
\frac{1}{\lambda f}(ux+vy)&=& \rho r (cos\phi cos\theta + sin\phi sin\theta)\nonumber \\
&=& \rho r cos(\phi - \theta)
\label{eq:refname23}
\end{eqnarray}

\begin{figure}[H]
\centering
\includegraphics[scale=0.5]{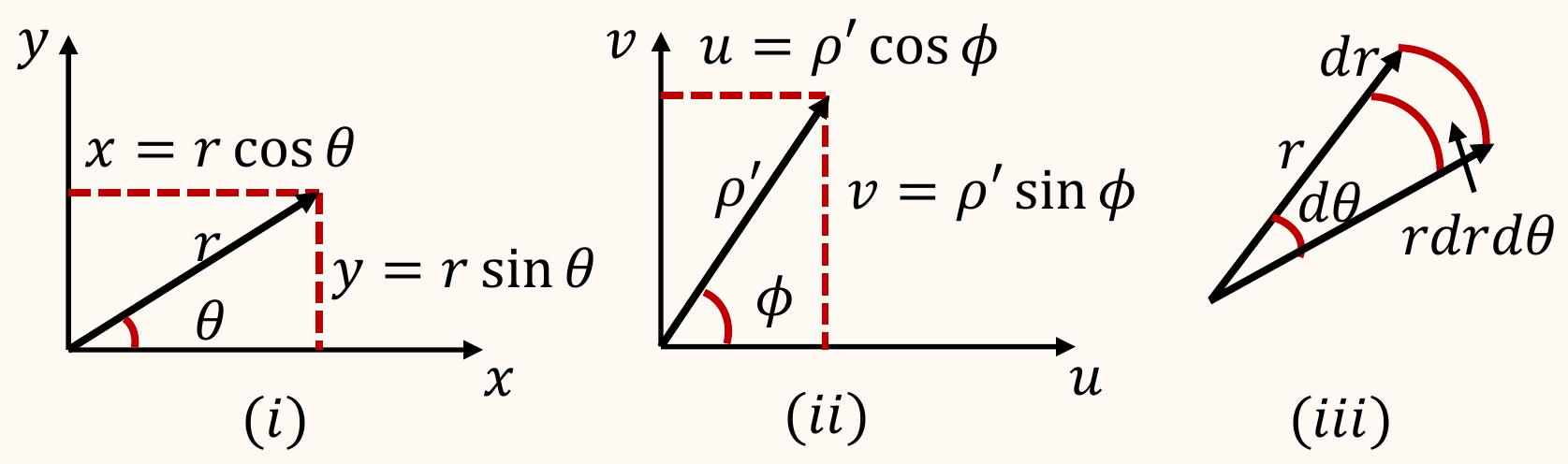}
\caption{Cartesian and polar coordinate representation of (i) $xy$ or $r \theta$ plane and (ii) $uv$ or $\rho \phi$ plane and (iii) representation of $dr$ and $d \theta$ in the $r \theta$ plane}
\label{fig19}
\end{figure}

The conversion of the Cartesian coordinates $x$, $y$, $u$ and $v$ into its polar form is shown in Fig. \ref{fig19}. The Cartesian coordinates of the plane of the lens ($x, y$) can be represented in terms of polar coordinates ($r, \theta$) as

\begin{eqnarray}
x = r cos \theta \nonumber \\
y = r sin \theta
\label{eq:refname24}
\end{eqnarray}

The  Cartesian coordinates ($u, v$) of the back focal plane of the lens can be represented in terms of polar coordinates ($\rho', \phi$) as

\begin{eqnarray}
u = \rho' cos \phi \nonumber \\
v = \rho' sin \phi
\label{eq:refname25}
\end{eqnarray}

The back focal plane of the lens can be numerically obtained by Fourier transforming the entity of the plane of the lens (i.e. $xy$ plane). In that case ($u, v$) coordinates will lie on the Fourier plane and $\rho' = \rho \lambda f$, where $\rho$ is the Fourier plane coordinate\cite{goodman2005introduction}. Thus the coordinates ($u, v$) can be represented in terms of polar coordinates ($\rho, \phi$) of the Fourier plane as

\begin{eqnarray}
u = \rho \lambda f cos \phi \nonumber \\
v = \rho \lambda f sin \phi
\label{eq:refname26}
\end{eqnarray}

which gives

\begin{eqnarray}
\frac{u}{\lambda f} = \rho cos \phi \nonumber \\
\frac{v}{\lambda f} = \rho sin \phi
\label{eq:refname27}
\end{eqnarray}

Now, from equation \ref{eq:refname24} and \ref{eq:refname27} we get

\begin{eqnarray}
\left(\frac{u}{\lambda f}\right)x + \left(\frac{v}{\lambda f}\right)y &=& \left(\rho cos\phi\right). \left(r cos\theta\right) + \left(\rho sin\phi\right). \left(r sin\theta\right) \nonumber \\
\frac{1}{\lambda f}(ux+vy)&=& \rho r (cos\phi cos\theta + sin\phi sin\theta)\nonumber \\
&=& \rho r cos(\phi - \theta)
\label{eq:refname28}
\end{eqnarray}

\end{multicols}
\end{document}